\input mssymb.tex

\def\inv{^{\raise.15ex\hbox{${
  \scriptscriptstyle -}$}\kern-.05em 1}}

\def\Dsl{\,\raise.15ex\hbox{$/$}\mkern-13.5mu D}
\def\dsl{\raise.15ex\hbox{$/$}\kern-.57em\hbox{$\partial$}}

\def\lspace{\ifx\answ\bigans{}\else\qquad\fi}
\def\del{\partial}
 \def\CC{\hbox{{$\cal C$}}}

\def\CR{\hbox{{$\cal R$}}} 
 \def\CV{\hbox{{$\cal V$}}}

\def\CO{\hbox{{$\cal O$}}}

\def\lform{\hbox{$\sqcup$}\llap{\hbox{$\sqcap$}}}
\def\darr#1{\raise1.5ex\hbox{$\leftrightarrow$}
\mkern-16.5mu #1}

\def\h{{{1\over2}}}

\def\INT{{\textstyle \int\kern-.642em\int}}

\def\R{{\Bbb R}}
\def\C{{\Bbb C}}
\def\Z{{\Bbb Z}}
\def\N{{\Bbb N}}

\def\eps{{\epsilon}}

\def\trace{{\rm Tr\, }}
\def\aut{{\rm Aut\, }}
\def\Rep{{\rm Rep\, }}

\def\cross{{\triangleright\!\!\!<}}
\def\cocross{{>\!\!\!\triangleleft}}
\def\small{\scriptstyle}
\def\tens{\mathop{\otimes}}

\def\la{{\triangleright}}\def\ra{{\triangleleft}}

\def\isom{{\cong}}

\def\Aut{{\rm Aut}}

\def\Ad{{\rm Ad}}

\def\ev{{\rm ev}}
\def\coev{{\rm coev}}

\def\id{{\rm id}}
\def\Deltaop{{\Delta^{\rm op}}}

\def\nquad{{\!\!\!\!\!\!}}
\def\nqquad{\nquad\nquad}\def\nqqquad{\nqquad\nquad}
\def\eqn#1#2{\begin{equation}#2\label{#1}\end{equation}}

\def\o{{}_{(1)}}\def\t{{}_{(2)}}\def\th{{}_{(3)}}

\def\calR{{\cal R}}\def\Ro{{\calR^{(1)}}}
\def\Rt{{\calR^{(2)}}}
\def\und#1{{\underline {#1}}}

\def\uo{{{}^{(1)}}}\def\ut{{{}^{(2)}}}

\def\new#1{\goodbreak\goodbreak\bigskip
\noindent{\bf #1}}

\def\text#1{\mbox{\rm #1}}
\def\note#1{}

\def\frac#1#2{{{#1\over#2}}}

\def\proof{\goodbreak\noindent{\bf Proof\quad}}
\def\endproof{{\ $\lform$}\bigskip }

\def\und#1{{\underline{#1}}}

\def\Bo{{{}_{\und{(1)}}}}\def\Bt{{{}_{\und{(2)}}}}
\def\vect{{\bf t}}\def\vecv{{\bf v}}
\def\vecu{{\bf u}}\def\vecx{{\bf x}}\def\vecp{{\bf p}}
\def\vecl{{\bf l}}

\def\<{\langle}
\def\>{\rangle}

\def\bmstat{R_{12}^{-1}\vecu'_1 R_{12}\vecu_2=\vecu_2 R_{12}^{-1}\vecu'_1
R_{12}}

\documentstyle[12pt,epsf]{article}
\newtheorem{lemma}{Lemma}[section]
\newtheorem{propos}[lemma]{Proposition}

\newtheorem{theorem}[lemma]{Theorem}

\newtheorem{corol}[lemma]{Corollary}

\begin{document}
\baselineskip 11pt
{\ } 
\begin{center}{\bf BEYOND SUPERSYMMETRY AND QUANTUM SYMMETRY}\\
{\bf (AN INTRODUCTION TO BRAIDED-GROUPS AND BRAIDED-MATRICES)}\footnote{1991
Mathematics Subject Classification 18D10, 18D35, 16W30, 57M25, 81R50, 17B37
$\qquad$
This paper is in final form and no version of it will be submitted for
publication elsewhere} \\
{\ }\\ {\ }\\
{\ }\\ {\small SHAHN MAJID}\footnote{SERC Fellow and Fellow of Pembroke
College, Cambridge}\\ {\ }\\
{\it Department of Applied Mathematics \& Theoretical Physics}\\ {\it
University of Cambridge, Cambridge CB3 9EW, U.K.}
\end{center}

\begin{quote}
\vskip 15pt
\centerline{\small ABSTRACT} \small
This is a systematic introduction for physicists to the theory of algebras and
groups with braid statistics, as developed over the last three years by the
author. There are braided lines, braided planes, braided matrices and braided
groups all in analogy with superlines, superplanes etc. The main idea is that
the bose-fermi $\pm1$ statistics between Grassmannn coordinates is now replaced
by a general braid statistics $\Psi$, typically given by a Yang-Baxter matrix
$R$. Most of the algebraic proofs are best done by drawing knot and tangle
diagrams, yet most constructions in
supersymmetry appear to generalise well. Particles of braid statistics exist
and can be expected to be described in this way. At the same time, we find many
applications to ordinary quantum group theory: how to make quantum-group
covariant (braided) tensor products and spin chains, action-angle variables for
quantum groups, vector addition on $q$-Minkowski space and a semidirect product
q-Poincar\'e group are among the main applications so far. Every quantum group
can be viewed as a braided group, so the theory contains quantum group theory
as well as supersymmetry. There also appears to be a rich theory of braided
geometry, more general than super-geometry and including aspects of quantum
geometry. Braided-derivations obey a braided-Leibniz rule and recover the usual
Jackson $q$-derivative as the 1-dimensional case.
\end{quote}
 \baselineskip 14pt

\section{Introduction}

There are certain physical situations in which it is known that particles of
braid-statistics exist. For example, the soliton sector in Yang-Mills theory in
2+1 dimensions with topological mass term\cite{Bal:che} and (related to this)
the theory of semions\cite{SorVol:ant}. Models with various anyonic statistics
are also known\cite{GMS:any}\cite{Wil:any} and there is a general theory that
any quantum field theory in two or three spacetime dimensions obeying
reasonable axioms with regard to locality, causality and structure of the
vacuum, has particles of braid statistics\cite{FRS:sup}\cite{Lon:ind}. Finally,
it has become apparent that q-deformation of physics (as in
q-regularization\cite{Ma:reg}) in any dimension
makes the particle statistics naturally braided\cite{Ma:poi}. The usual
spin-statistics theorem fails  in the lower-dimensions for
familiar reasons (in 1+1 dimensions because the complement of a light-cone is
not connected) and in the q-deformed case because the space-time itself is
non-commutative.

The existence of these new particles with braid statistics then motivates the
development of an algebraic framework with braided-commuting variables, just as
the existence of particles with bose-fermi statistics in 3+1 dimensions
motivated the development of the algebraic framework of super-algebras based on
Grassmann variables. Here I want to give a systematic account of such an
algebraic framework for braided particles, {\em braided algebra}, introduced
over a series of papers spanning the last three years. The main works are
\cite{Ma:bra}\cite{Ma:exa}\cite{Ma:lin}\cite{Ma:csta} (for physicists) and
\cite{Ma:eul}\cite{Ma:bg}\cite{Ma:tra}\cite{Ma:bos}\cite{Ma:cat} (for
mathematicians), while the remaining 20 or so are applications (many of them
with collaborators). Since the subject is now beginning to be appreciated by
physicists, perhaps the time is right for such a review. It will be aimed at
theoretical physicists, although Chapter~4 should be of use to mathematicians
also. An earlier review for physicists appeared in \cite{Ma:sta}.

We will describe a variety of braided lines, braided planes, braided matrices
etc, as well as braided-differential calculus. The basis of this generalization
is to replace the $\pm 1$ phases encountered for Grassmann variables by a more
general phase\cite{Ma:any}\cite{Ma:csta} or more generally by an $R$-matrix (a
matrix obeying the celebrated quantum Yang-Baxter equations). These are the
generalised braid statistics $\Psi$ and play the role of usual transposition or
super-transposition. Our work differs in one fundamental respect from previous
attempts to generalise super-symmetry in this way (notably the theory of
colour-Lie algebras \cite{Sch:gen} where there is a phase given by a skew
bicharacter on an group, and the theory of S-Lie
algebras\cite{Gur:alg}\cite{Man:non} and others, where the phase is generalised
by a triangular $R$-matrix.) In all these works the $R$-matrix or colouring
bicharacter is such that the resulting $\Psi$ was symmetric in the sense
$\Psi^2=\id$. This is a natural condition for any analogue of transposition or
super-transposition but meant the resulting theory was too close to the usual
theory of supersymmetry. In particular, there is no braiding in the picture at
all. The main task of our braided theory was to show how to relax this symmetry
condition and hence reach the case where $\Psi\ne\Psi^{-1}$, and it is this
aspect which leads to many unusual features in the theory to be described. As a
result, $\Psi$ and $\Psi^{-1}$ are more properly represented by mutually
inverse braid crossings and indeed, most of the algebraic manipulations are now
best done by means of braid and tangle diagrams.
The general setting is explained in Chapter~3.

In principle, this line of development leading to braided geometry has nothing
whatever to do with quantum inverse scattering (QISM) and quantum
groups\cite{Dri}\cite{Jim:dif}\cite{FRT:lie} even though these are also
obtained from the same data. This is because the same phase factors or
$R$-matrices can be used on the one hand to define associative non-commutative
algebras (the philosophy of non-commutative geometry\cite{Con:alg}) or on the
other hand to define non-commutative statistics (the philosophy of
super-geometry and its generalizations). This is depicted in Figure~1.
Not only are the mathematical ideas quite orthogonal but the physical
distinction is also clear. Thus, in quantum field theories which are
characterised by a  `functor' of some form (in which the super-selections
sectors are mapped to Hilbert spaces), such as any conformal field theory or
topological quantum field theory, there are general arguments\cite{Ma:int} that
there is always some kind of weak quantum group of internal symmetries. By
contrast, in a general non-topological 1+1 or 2+1 quantum field theory there is
no such functor to Hilbert spaces, hence there is no reason to find any kind of
quantum group. Instead, we proved a generalised reconstruction theorem and
introduced the notion of braided group for the resulting
structure\cite{Ma:rec}\cite{Ma:bra}. We proved that if $\CC$ is any braided
collection of objects (braided category) then there is an underlying braided
group $\aut(\CC)$.

\begin{figure}
\vskip .3in
\qquad\epsfxsize=12cm\epsfbox{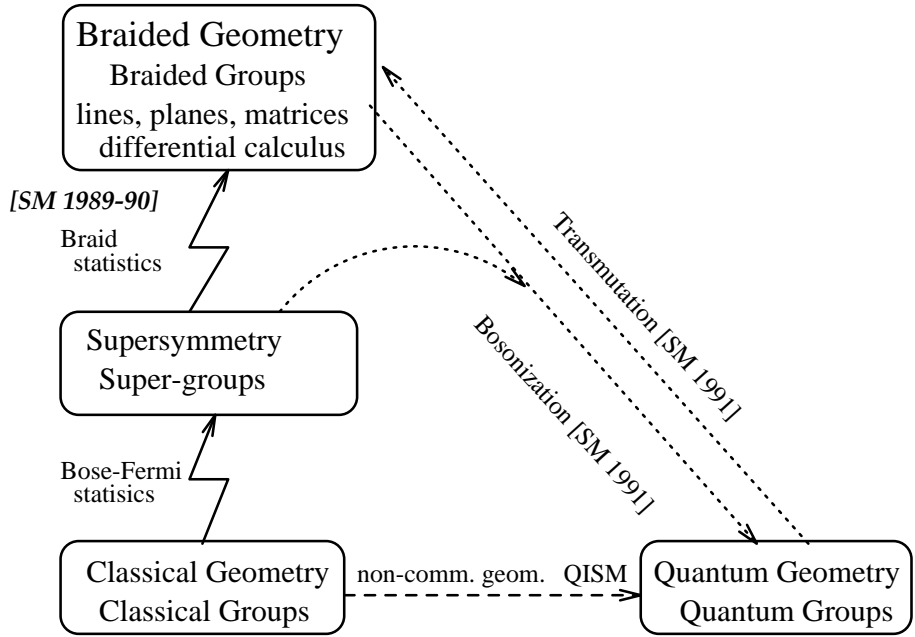}
\caption{Braided groups extend the philosophy of super-symmetry}
\end{figure}

The same distinction holds mathematically. Thus, any $R$-matrix generates a
braided category $\CC$. There is a functor
and using it one reconstructs by standard arguments the quantum matrices $A(R)$
of \cite{FRT:lie}. By contrast, proceeding without the functor one reconstructs
$\aut(\CC)=B(R)$, the braided-matrices. Thus quantum and braided matrices are
cousins but have quite different origins and philosophy. Chapter~2 begins with
an introduction to braided matrices with emphasis on this contrast. We explain
what exactly is a braided matrix and in what sense it generalises a
super-matrix. The chapter does not assume a background in quantum groups.

Although having a very different origin, it is natural to compare and contrast
braided groups and quantum groups. Just as simple Lie groups have quantum group
analogs, they also have braided-group analogues. We have already mentioned that
every $R$-matrix leads to a braided matrix group just as it leads to a quantum
matrix group. As a matter of fact, every strict quantum group (equipped with a
universal $R$-matrix or its dual) has a braided-group analogue. This process of
conversion of a strict quantum group into a braided group is a process we call
{\em transmutation}\cite{Ma:tra}. There is also an `adjoint' process that
converts any braided-group of a certain type into an ordinary (bosonic) quantum
group\cite{Ma:bos}. These form the topic of Chapter~5. These mathematical
theorems mean that braided groups, while more general, also provide a tool for
working with quantum groups. Many quantum group constructions are much simpler
when viewed as corresponding braided group constructions. The general reason is
that braided groups, like super-groups, are in some sense classical (not
quantum) objects, albeit in a modified sense. Hence super-geometry and braided
geometry are much closer to classical geometry than the quantum case. Some
applications to quantum groups are

\begin{itemize}

\item the {\em braided tensor product} -- how to tensor product quantum-group
covariant systems to obtain another such.  The construction is by analogy with
the super or $Z_2$-graded tensor product (The notions of statistics and
group-covariance are unified in the notion of quantum symmetry). This is
fundamental to the very definition of braided matrices and braided groups (and
we introduced it for this purpose).

\item quantum-covariant spin chains -- obtained by interated braided tensor
products. Examples include a system of n-braided harmonic oscillators
\cite{BasMa:bra}\cite{BasMa:unb} (w/ W.K. Baskerville) and exchange algebras in
2-D quantum gravity \cite{Ma:inf}.

\item the {\em quantum Fourier transform} on factorizable quantum groups
\cite{LyuMa:bra}\cite{LyuMa:fou} (w/ Lyubashenko). When quantum groups are
viewed in our braided setting they appear more like $\R^n$ than anything else.

\item Using braided matrices one can construct a mutually commuting set of
`angle' variables $\alpha_k$ for any quantum group $A(R)$. One also recovers
Casimirs etc as {\em bosonic} elements from the point of view of the braided
group.

\item quantum differential calculus -- a general construction for bicovariant
quantum differential calculus that includes all known examples \cite{BrzMa:bic}
(w/ T. Brzezinski). Any function of the $\alpha_k$ defines a calculus, so we
have a new function or field in q-deformed physics corresponding to the choice
of differential calculus.

\item semidirect product structure of Drinfeld's quantum double $D(H)$ -- which
then leads to its interpretation as a quantum algebra of observables on a
braided group quantized by Mackey's method\cite{Ma:mec}. In the factorizable
case we also have $D(H)\isom H\cocross H$ by the quantum adjoint
action\cite{Ma:skl}. For example, $D(U_q(g))$ is generated by two copies of
$U_q(g)$ with relations given by the quantum adjoint action (also computed in
\cite{Ma:skl}) and leading to a kind of quantum Lie bracket.

\item $U_q(g)$ as a kind of deformation-quantization -- \cite{GurMa:bra} (w/ D.
Gurevich).

\item Spectrum generating quantum groups -- some braided groups, including all
super groups, can be {\em bosonized} to equivalent quantum groups. Thus
statistical non-commutativity can be swapped for quantum non-commutativity.
This is physically more appropriate in some
situations\cite{MacMa:spe}\cite{MacMa:str} (w/ A.J. Macfarlane).

\item Superization and Anyonization -- some quantum groups are more naturally
super quantum groups or anyonic-quantum groups. For example, the super-like
aspects of the  non-standard quantum group of Wu and Jing, are understood by
showing that when superized it is the super-quantum group $GL_q(1|1)$. The
usual and graded FRT constructions are related like this\cite{MaPla:uni} (w/
M-J. Rodriguez-Plaza).
\end{itemize}

Some of these are outlined in Chapter~6. Thus braided groups and braided
matrices are a useful tool for obtaining results about quantum groups. What
about situations where we genuinely need braided groups and braided matrices?
Some results at the time of writing are

\begin{itemize}

\item group-like structure on the degenerate Sklyanin algebra -- many authors
tried to a find a quantum group here and failed. It turns out to be the braided
$2\times 2$ matrices $BM_q(2)$ \cite{Ma:skl}. This is important because it
means that one can tensor product representations of the degenerate Sklyanin
algebra in a way familiar for groups and quantum groups (but remembering the
braid statistics).

\item group-like structure on quantum homogeneous spaces $G/H$ -- even in the
normal case, if $H\subset G$ are two quantum groups, the quotient is not a
group or quantum group but a braided group. This supplies examples of quantum
group principal bundles\cite{BrzMa:gau}\cite{BrzMa:mon} (w/ T. Brzezinski).

\item group-like structure for $q$-Minkowski space -- many authors tried to
find a quantum group here, but failed. It turns out to be a braided group
$\R^{1,3}_q$ (of vector addition). As a consequence, we also obtain the
$q$-Poincare quantum group of Schlieker et al as the semidirect product
$\widetilde{SO_q(1,3)}\cross \R^{1,3}_q$\cite{Ma:poi}.

\item braided-differential calculus -- operators $\del^i$ generate the braided
addition law on ${\R^{1,3}_q}$ and obey a braided-Leibniz rule. The usual
Jackson derivative familiar in q-analysis is the 1-dimensional case and is
thereby generalized to $n$-dimensions by means of an $R$-matrix.

\end{itemize}

Some of these are outlined in Chapter~7. This completes our outline of the
theory and of the paper. Perhaps the simplest examples other than the general
braided matrices (which we begin with in the next chapter) are the
braided-line, either anyonic as in \cite{Ma:any} or $\C$-statistical as in
\cite{Ma:csta} and as outlined at the start of Chapter~7.3.

\new{Historical Note} The reconstruction theorem leading to braided groups
$\Aut(\CC)$ was circulated in preprint form in 1989 and was finally published
(after two rejections) in \cite{Ma:rec}. The resulting notion of braided groups
was presented in May~1990 at the conference on {\em Common Trends in
Mathematics and Quantum Field Theory}, Kyoto. The result was also presented at
the {\em XIX DGM}, Rapallo, June 1990\cite{Ma:som} and at the {\em Euler
Institute Programme on Quantum Groups}, Leningrad, October~1990. The explicit
formulae for the braided matrices $B(R)$ (\ref{bra-com})-(\ref{bra-mat}) and
the example $BSL_q(2)$ were obtained at the end of 1990 and appeared in the
proceedings of the Leningrad conference\cite{Ma:eul}, edited by P.P. Kulish
(whom I thank for proof-reading the manuscript). At this time the equations
were written with indices and the $R$'s on one side rather than in the compact
form. Further examples were presented in December 1990 at the conference on
{\em Quantum Probability and Related Topics}, Delhi, including the braided
matrices $BM_q(1|1)$ associated to the $R$-matrix for the Alexander-Conway knot
polynomial, and published in\cite{Ma:sta}. The main mathematical works
\cite{Ma:bg}\cite{Ma:tra}\cite{Ma:bos} were circulated in preprint form in late
1990 -- early 1991, with the key part presented at the {\em Biannual Meeting of
the American Maths Society}, San Francisco, January 1991.
Thus the main part of the work presented in this review dates from the period
1989 -- early 1991. The latter half of 1991 and the subsequent time has been
devoted to developing applications of the theory, some of it with
collaborators. For example, the work with V.~Lyubashenko \cite{LyuMa:bra} made
contact with some of his independent constructions for braided categories with
left and right duals\cite{Lyu:mod}, while the work with D.
Gurevich\cite{GurMa:bra} generalised some of his work on $S$-quantization in
\cite{Gur:alg} to the braided case. The compact notation
(\ref{bra-com})-(\ref{bra-mat}) for the braided matrices was explicitly
introduced in \cite{Ma:skl}\cite{Ma:lin} circulated in January and February
1992 respectively, where we recognized the first of these as equations of
interest in another context in QISM, as for example in \cite{ResSem:cen}. The
braided-matrix property and braid-statistics $\bmstat$ however, arose only in
the theory of braided matrices and not in QISM.

It is hoped by means of this chronology to make clear the historical origin of
the theory of braided groups and braided matrices. In particular, the braided
matrix property expressed in (\ref{bra-com})-(\ref{bra-mat}) did not arise in
any way in connection with the reflection equations in \cite{Skl:bou} as
sometimes suggested. Perhaps a significant connection can be established in the
future but so far no results about braided matrices have been obtained from
this point of view beyond those already published a year ago in
\cite{Ma:bra}\cite{Ma:exa}. Covariance (\ref{Ad-coact-mat}) of the
braided-commutativity equation (\ref{bra-com}), its braid-diagrammatic form
(\ref{comtens}), the study of the resulting quadratic algebras for non-standard
$R$-matrices and finally the braided-matrix property were some of the first
main results in the theory of braided groups and appeared in some form in
\cite{Ma:bra}\cite{Ma:exa}\cite{Ma:eul} and numerous preprints and other
publications dating from 1989-1990.

The braided-matrix results in the compact notation (and applications) have been
presented at various conferences, most recently in a series of four talks here
at the Nankai Maths Institute. I would like to thank the organisers, especially
M-L. Ge for inviting me give two of these talks (at this Workshop and at the
{\em XXI DGM} immediately preceding it) and P.P. Kulish and R. Sasaki for
encouraging me to continue my physicists account of braided groups and braided
matrices in two further unofficial seminars here in the Nankai Institute. I
would like to thank all concerned for a very enjoyable visit to the Nankai
institute.

\section{Braided Linear Algebra}

We begin by explaining what is a matrix from the point of view of algebraic
geometry. In fact, this point of view is entirely familiar to physicists in the
context of classical mechanics. The main idea is that instead of working
directly with matrices, we can work with the algebra of functions $C(M_n)$
on the matrices. This is generated by the co-ordinate functions $t^i{}_j\in
C(M_n)$, thought of as abstract `matrix-entry observables'  with value
$t^i{}_j(M)=M^i{}_j$ on any matrix $M$. Because they commute, they can all
simultaneously have values of this form. Indeed, any actual matrix $M$
determines a state (a linear functional) on the algebra $C(M_n)$ viewed as a
classical algebra of observables of a classical system. In the corresponding
representation, the mutually commuting observables $t^i{}_j$ have the
simultaneous values $M^i{}_j$.

In this slightly unfamiliar language, the fact that we can multiply matrices
expresses itself as follows. If $\vect$ and $\vect'$ are the generators of two
independent (commuting) copies of $C(M_n)$, then $\vect\vect'$ is a realisation
of $C(M_n)$ in $C(M_n)\tens C(M_n)$. It is the collection of observables on the
joint system whose value is the matrix product of the values in  each system,
$t''^i{}_j(M\tens N)= t^i{}_k(M) t'^k{}_j(N)= M^i{}_k N^k{}_l$. In terms of the
algebraic properties of the $t^i{}_j,t'^i{}_j$ alone, the key property is
\eqn{com-mat}{[t'^i{}_j,t^k{}_l]=0,\quad t''^i{}_j=t^i{}_kt'^k{}_j\
\Rightarrow\  [t''^i{}_j,t''^k{}_l]=0}
so that $\vect''$ realises the same commutative algebra $C(M_n)$ in the joint
system.

There are two quite distinct directions in which to generalise this notion, in
both of which we drop the actual matrices $M,N$ and work with the algebraic
structure of the $t^i{}_j$ directly and in a generalized form. The first(and
more topical) direction is that of a {\em quantum matrix}. Here we allow the
algebra $C(M_n)$ to be replaced by a non-commutative algebra $A(R)$ (say),
still with generators $t^i{}_j$. Now these no-longer commute. Typically, their
quantum-commutation relations are of the form\cite{FRT:lie}
\eqn{FRT}{R^i{}_m{}^k{}_n t^m{}_j t^n{}_l=t^k{}_n t^i{}_n
R^m{}_j{}^n{}_l,\qquad {\rm i.e.\ }\quad R\vect_1\vect_2=\vect_2\vect_1 R}
where the second expression is a compact notation where the numerical suffices
refer to the position in the matrix tensor product, and $R\in M_n\tens M_n$
obeys the QYBE $R_{12}R_{13}R_{23}=R_{23}R_{13}R_{12}$ in the same compact
notation. Thus $A(R)$ is some non-commutative `quantized' version of $C(M_n)$
but the generators $t^i{}_j$ can still be multiplied as before, in a way that
would correspond (if the $t^i{}_j$, $t'^i{}_j$ had definite values) to matrix
multiplication. Thus the analog of (\ref{com-mat}) is
\eqn{q-mat}{[t'^i{}_j,t^k{}_l]=0,\quad t''^i{}_j=t^i{}_kt'^k{}_j\ \Rightarrow\
R\vect''_1\vect''_2=\vect''_2\vect''_1R.}
This says that the matrix product $\vect''$ obeys the same quantum-commutation
relations and so realises a copy of $A(R)$ in the joint system $A(R)\tens A(R)$
generated by the two independent (commuting) quantum matrices $\vect,\vect'$.

A second (and even more well-known) direction is that of a {\em super-matrix}.
Here the philosophy is quite different in that we keep the algebra
`commutative' (not quantized) but in the generalised sense of being
super-commutative. Thus the generators have a grading such as $|t^i{}_j|=i+j$
mod $2$, and obey the relations
\eqn{sup-com}{t^i{}_j t^k{}_l=(-1)^{|t^i{}_j||t^k{}_l|}t^k{}_l t^i{}_j.}
The ability to multiply super-matrices appears now as the following: if
$t'^k{}_l$ is another independent copy of the super-matrix, independent now
meaning that it has super-statistics with $t^i{}_j$, then the product is also a
super-matrix. Thus
\eqn{sup-mat}{t'^i{}_j t^k{}_l=(-1)^{|t'^i{}_j||t^k{}_l|}t^k{}_l
t'^i{}_j,\quad t''^i{}_j=t^i{}_kt'^k{}_j\ \Rightarrow\ t''^i{}_j
t''^k{}_l=(-1)^{|t''^i{}_j||t''^k{}_l|}t''^k{}_l t''^i{}_j.}
Thus the idea of a super-matrix is somewhat different from that of a quantum
matrix. In the super case, the algebra remains `commutative' and hence
classical but in a modified sense appropriate to a modified exchange law or
statistics between independent copies.

The idea of a {\em braided matrix} as introduced in \cite{Ma:exa}\cite{Ma:eul}
is to use the same data $R$ obeying the QYBE but to use it according to the
philosophy of the super case rather than the more conventional quantum one.
Let us call the braided-matrix generators $u^i{}_j$ to distinguish them from
the above. They generate an algebra $B(R)$
with {\em braided-commutativity relations}
\eqn{bra-com}{R^k{}_a{}^i{}_b u^b{}_c R^c{}_j{}^a{}_d u^d{}_l=u^k{}_a
R^a{}_b{}^i{}_c u^c{}_d R^d{}_j{}^b{}_l\quad {\rm i.e.}\quad R_{21}\vecu_1
R_{12}\vecu_2=\vecu_2 R_{21}\vecu_1 R_{12}}
and if $\vecu'$ is another independent braided matrix obeying the same
relations and having certain braid-statistics with $\vecu$ then the product
$\vecu\vecu'$ is also a braided-matrix,
\eqn{bra-mat}{R_{12}^{-1}\vecu'_1 R_{12}\vecu_2=\vecu_2 R_{12}^{-1}\vecu'_1
R_{12},\quad u''^i{}_k=u^i{}_ku'^k{}_j\ \Rightarrow\ R_{21}\vecu''_1
R_{12}\vecu''_2=\vecu''_2 R_{21}\vecu''_1 R_{12}.}
The relations between $\vecu,\vecu'$ are the {\em braid-statistics relations}
between two independent identical braided matrices\cite{Ma:exa}. Given these,
the proof of (\ref{bra-mat}) in the present compact notation\cite{Ma:lin} is
$R_{21}\vecu''_1 R_{12}\vecu''_2=R_{21}\vecu_1\vecu'_1
R\vecu_2\vecu'_2=R_{21}\vecu_1 R(R^{-1}\vecu'_1
R\vecu_2)\vecu'_2=(R_{21}\vecu_1 R\vecu_2) R^{-1}R_{21}^{-1}(R_{21}\vecu'_1
R\vecu'_2)= \vecu_2 R_{21}(\vecu_1 R_{21}^{-1}\vecu'_2 R_{21})\vecu'_1
R=\vecu_2 R_{21}R_{21}^{-1}\vecu'_2 R_{21}\vecu_1\vecu'_1 R$
$= \vecu''_2 R_{21}\vecu''_1 R$ as required. In each expression, the brackets
indicate how to apply the relevant relation to obtain the next expression.

Note that if $R_{21}=R_{12}^{-1}$ (the so-called triangular or unbraided case)
the braid-statistics relations and the braided-commutativity relations
(\ref{bra-com}) coincide. This is the case for super-matrices (which fit into
this framework for suitable $R$) but in the general braided case, the notion of
braided-commutativity and braid-statistics are slightly different. This was one
of the key obstacles to the braided case that was solved in \cite{Ma:exa}.

For one of the simplest non-trivial examples we take $R$ to be the standard
$SL_q(2)$ $R$-matrix, giving the braided matrices
$BM_q(2)$\cite{Ma:exa}\cite{Ma:eul}. Denoting the matrix entries as
$\vecu=\pmatrix{a& b\cr c&d}$, the braided-commutativity relations become
\eqn{bm2a}{ba=q^2ab,\qquad ca=q^{-2}ac,\qquad da=ad,\qquad
bc=cb+(1-q^{-2})a(d-a)}
\eqn{bm2b}{db=bd+(1-q^{-2})ab,\qquad cd=dc+(1-q^{-2})ca}
while the braid-statistics relations between two independent copies comes out
as
\eqn{bm2stat}{\begin{array}{rlll}
\nqquad&a'  a=a a'+(1-q^2)b c' &a'  b=b  a'& \nqqquad\nqquad a'  c=c
a'+(1-q^2)(d-a)  c'\\
\nqquad&a'  d=d  a'+(1-q^{-2})b  c' &\nquad b'  a=a  b'+(1-q^2)b  (d'- a') &b'
b=q^2b  b' \\
\nqquad&b'  c=q^{-2}c  b'+(1+q^2)(1-q^{-2})^2b  c'-
(1-q^{-2})(d-a) (d'-a')\nqqquad\nqqquad\nqqquad\nqqquad \nqqquad \nqqquad
\nqqquad &\ &\ \\
\nqquad&b'  d=d  b'+(1-q^{-2})b (d'-a') &\qquad c'  a=a  c' &c'  b=q^{-2}b  c'
\\
\nqquad&c'  c=q^2c  c' &\nquad c'  d=d  c' &\nqqquad d'  a=ad'+(1-q^{-2})b  c'
\\
\nqquad&d'  b=b  d'&\nquad\nqqquad\nqqquad d'  c=c  d'+(1-q^{-2})(d-a)
c'&\nqquad\nqqquad d'  d=d  d'-q^{-2}(1-q^{-2})b  c'.
\end{array}}
In this example, as $q\to 1$ the algebra becomes commutative and the statistics
also become commutative, so we return to the case (\ref{com-mat}).

For another choice of $R$-matrix, for example the non-standard $R$-matrix
associated to the Alexander-Conway knot polynomial, we arrive at the braided
matrices $BM_q(1|1)$ as computed in \cite[Sec. 3]{Ma:exa}. Its
braided-commutativity relations are
\eqn{bm1|1a}{b^2=0,\qquad c^2=0,\qquad d-a\  {\rm central}, }
\eqn{bm1|1b}{ab=q^{-2}ba,\qquad ac=q^2 ca,\qquad bc=-q^{2}cb+(1-q^2)(d-a)a}
and its braid-statistics relations are $d-a$ bosonic and
\eqn{bm1|1stat}{\begin{array}{rll}
 &a'a=a a' + (1-q^2) bc'\qquad\qquad&b'b=-bb'\\
&a'b=ba'\qquad\qquad\qquad\qquad\qquad&b'c=-c b'- (1-q^2)(d-a)(d'-a')\\
&b'a=ab'+(1-q^2)b(d'-a')\quad
&c'b=-bc'\\
&a'c=ca'+ (1-q^2)(d-a)c'\quad&c'c=-cc'\\
&c'a=ac'.\qquad\qquad\qquad\qquad\qquad&
\end{array}}
Here, in the limit $q\to 1$ the commutativity and statistics relations become
exactly the ones for the super-matrices $M(1|1)$ (with $b,c$ odd and $a,d$
even) as in the (\ref{sup-com})-(\ref{sup-mat}). Thus the notion of braided
matrices really generalises both ordinary and super-matrices. In the general
case the statistics are not merely a phase as in the super case, but given by a
linear combination via an $R$-matrix. The braided-matrix property means that
\eqn{bmat-mult}{\pmatrix{a''&b''\cr c''& d''}=\pmatrix{a&b\cr
c&d}\pmatrix{a'&b'\cr c'&d'}}
also obeys the braided-matrix relations {\em provided} we remember the
braid-statistics.

There are also braided-traces and braided-determinants in analogy with the
super-case. Let $\widetilde R=((R^{t_2})^{-1})^{t_2}$ where $t_2$ is
transposition in the second matrix factor of $M_n\tens M_n$ and we assume that
the relevant inverse here exists. Let $\vartheta^i{}_j=\widetilde
R^i{}_k{}^k{}_j$ then\cite{Ma:lin}
\eqn{bos}{c_k=\trace \vartheta \vecu^k\ \Rightarrow  \quad \vecu'
c_k=c_k\vecu',\qquad \vecu c_k=c_k\vecu}
so that the $c_k$ are bosonic and central elements of the braided matrix
algebra $B(R)$. The element $c_1$ is the braided-trace while products of the
$c_k$ can be used to define the braided-determinant. The braided-determinant
for the $SL_q(2)$ $R$-matrix is
\eqn{bm2b-det}{BDET(\vecu)=ad-q^2 cb.}
Setting this to $1$ in $BM_q(2)$ gives the braided group
$BSL_q(2)$\cite{Ma:exa}\cite{Ma:eul}.

In super-symmetry of course, we have not only super-matrices but super-vectors
and super-covectors with a linear addition law. Thus if we denote as usual the
Grassmann variables by $\theta_i$ with degree $|\theta_i|=i-1$ mod 2 say, we
have super-commutativity

\eqn{sup-covec-com}{\theta_i\theta_j=(-1)^{|\theta_i||\theta_j|}
\theta_j\theta_i}
and if $\theta'$ is another copy of the super-plane with super-statistics
relative to $\theta$, we can add them. Thus
\eqn{sup-covec}{\theta'_i\theta_j=(-1)^{|\theta'_i||\theta_j|}\theta_j
\theta'_i,\quad \theta''_i=\theta_i+\theta'_i\ \Rightarrow\
\theta''_i\theta''_j=(-1)^{|\theta''_i||\theta''_j|}\theta''_j\theta''_i.}
In the usual case these $\theta_i$ would be the commuting co-ordinate functions
generating $C(\R^n)$ and their linear addition would express linear addition on
$\R^n$. In the super-case we work abstractly with the generators $\theta_i$,
just as we did for the $t^i{}_j$ above, there being no underlying actual points
any more.

Here our notion of braided linear algebra strikes an important success over its
rival notion of quantum-linear algebra: quantum planes (such as the
much-discussed quantum planes $\C^{2|0}_q,\C^{0|2}_q$) can be defined with
non-commutation relations among their generators, but these are {\em not}
preserved under linear addition. Hence, there is no real quantum linear algebra
(with linear addition of quantum vectors). By contrast, generalising the super
case, we do have braided-vectors and braided-covectors. Denoting the covector
generators $x_i$ we use them to generate a braided-commutative algebra
$V^*(R')$ of `co-ordinate' functions with braided-commutation relations
\eqn{bra-covec-com}{x_j x_l=x_n x_m R'^m{}_j{}^n{}_l,\quad {\rm i.e.}\quad
\vecx_1\vecx_2=\vecx_2\vecx_1R'_{12}.}
Here $R'$ is built from $P$ (the usual permutation matrix) and $R$, and is
characterised by the equation
\eqn{R'}{(PR+1)(PR'-1)=0.}
For example, there is an $R'$ for each non-zero eigenvalue $\lambda_i$ in the
functional equation $\prod_i(PR-\lambda_i)=0$. First one should rescale $R$ by
dividing by $-\lambda_i$, and then $R'=P+\mu P\prod_{j\ne
i}(PR+(\lambda_j/\lambda_i))$. Here $\mu$ is an arbitrary non-zero constant
which does not change the algebra and which can be chosen so that $R'$ is
invertible. The linear addition of braided-covectors means that if $\vecx'$ is
another copy with braid-statistics relative to $\vecx$ then $\vecx+\vecx'$ is
also a realization of the same algebra,
\eqn{bra-covec}{\vecx'_1\vecx_2=\vecx_2\vecx'_1R_{12},\quad
\vecx''=\vecx+\vecx'\ \Rightarrow\
\vecx''_1\vecx''_2=\vecx''_2\vecx''_1R'_{12}.}
The proof is
$(\vecx_1+\vecx'_1)(\vecx_2+\vecx'_2)=\vecx_1\vecx_2+\vecx'_1\vecx_2+\vecx_1
\vecx'_2+\vecx'_1\vecx'_2=\vecx_2\vecx_1R'_{12}+\vecx_1\vecx'_2 (PR_{12}+1)
+\vecx'_2\vecx'_1R'_{12}$ while $(\vecx_2+\vecx'_2)(\vecx_1+\vecx'_1)R'_{12}$
has the same outer terms and the cross terms $\vecx'_2\vecx_1R'_{12}+\vecx_2
\vecx'_1 R'_{12}=\vecx_1\vecx'_2(R_{21}+P)R'$. These are equal since
$PR+1=PRPR'+PR'$ from (\ref{R'}).

Note that if $R$ is a Hecke symmetry in the sense $(PR+1)(PR-q^2)=0$ (such as
for all the $SL_q(n)$ $R$-matrices) we can take $R'=q^{-2}R$. In this case, our
braided-covectors reduce to the usual Zamolodchikov algebra.
For example, the $SL_q(2)$ $R$-matrix gives for $V^*(R)$ the usual quantum
planes $\C_q^{2|0}$ or $\C_q^{0|2}$ according to the chosen eigenvalue. Thus
$\C_q^{2|0}$ has generators $x,y$ and braided-commutativity relations
$xy=q^{-1}yx$. The corresponding braid-statistics relations are\cite{Ma:lin}
\eqn{bra-2|0-stat}{x'x=q^2 xx',\  x'y=q yx',\  y'y=q^2 yy',\ y'x=q
xy'+(q^2-1)yx'.}
Similarly, $\C_q^{0|2}$ has generators $\theta,\eta$ say, and
braided-commutativity relations $\theta^2=0, \eta^2=0, \theta\eta=-q\eta\theta$
and braid-statistics relations
\eqn{bra-0|2-stat}{\theta'\theta=- \theta\theta',\
\theta'\eta=-q^{-1}\eta\theta',\  \eta'\eta=- \eta\eta',\ \eta'\theta=-q^{-1}
\theta\eta'+(q^{-2}-1)\eta\theta'.}
In the limit $q\to 1$ we obtain exactly the usual $\C^{0|2}$ plane with
$\theta,\eta$ fermionic.

By contrast, the Alexander-Conway $R$-matrix mentioned above gives two algebras
$\C_q^{1|1}$ according to the eigenvalue. One has generators $x,\theta$ say,
with braided-commutativity relations $x\theta=q^{-1}\theta x, \theta^2=0$ and
braid-statistics relations\cite{Ma:lin}
\eqn{bra-1|1stat}{x'x=q^2 xx',\  x'\theta=q \theta x',\
\theta'\theta=-\theta\theta',\ \theta'x=q x\theta'+(q^2-1)\theta x'.}
In the limit $q\to 1$ we obtain exactly the $\C^{1|1}$ super-plane with
$\theta$ fermionic and $x$ bosonic. The other eigenvalue is similar with the
first co-ordinate fermionic. These examples are all of Hecke type.

Other $R$-matrices give more complicated algebras, for example the $SO_q(1,3)$
$R$-matrix gives the $q$-Minkowski space algebra as in\cite{CWSSW:lor} but now
equipped with a braided addition law. The braided-vectors $V(R')$ are similar,
with generators $v^i$ and relations $\vecv_1\vecv_2=R'\vecv_2\vecv_1$ and a
braided-addition law. Thus,
\eqn{bra-vec}{\vecv'_1\vecv_2=R_{12}\vecv_2\vecv'_1,\quad \vecv''=\vecv+\vecv'\
\Rightarrow\ \vecv''_1\vecv''_2=R'_{12}\vecv''_2\vecv''_1.}
The proof is similar to that above. There are also other variants of
$V(R'),V^*(R'),B(R)$ according to other conventions (our present conventions
are right-handed as we will see later).

This describes braided-matrices, braided-vectors and braided-covectors in
isolation. However, they are all part of a single unified braided-linear
algebra. Thus just as in super-symmetry, where all algebras have
super-statistics relative to each other, so there are braid-statistics between
the braided matrices, braided-vectors and covectors\cite[Lemma~3.4]{Ma:lin}
\eqn{bra-co-vec-stat}{\vecx'_1 R\vecv_2=\vecv_2\vecx'_1,\quad
\vecv'_1\vecx_2=\vecx_2R^{-1}\vecv'_1}
\eqn{bra-mat-covec-stat}{\vecu'_1\vecx_2=\vecx_2 R^{-1}\vecu'_1 R,\quad
\vecx'_1R\vecu_2R^{-1}=\vecu_2\vecx'_1}
\eqn{bra-mat-vec-stat}{\vecv'_1\vecu_2=R\vecu_2 R^{-1}\vecv'_1,\quad
R^{-1}\vecu'_1 R\vecv_2=\vecv_2\vecu'_1.}

For example, remembering such braid-statistics, we can act on a braided
covector $\vecx$ by a braided-matrix $\vecu$ in the sense that
$\vecx'=\vecx\vecu'$ obey the same relations
$\vecx'_1\vecx'_2=\vecx_1\vecu'_1\vecx_2\vecu'_2= \vecx_1\vecx_2R^{-1}\vecu'_1
R\vecu'_2=\vecx_2 \vecx_1 R'R^{-1}\vecu'_1 R\vecu'_2=\vecx_2\vecx_1
R_{21}^{-1}\vecu_2' R_{21}\vecu_1' R'=\vecx_2\vecu_2'\vecx_1\vecu_1'R'$ as
required. Here we used that $R'$ is some function $R'=Pf(PR)$ (see above) and
\eqn{PR-bracom}{(PR)R^{-1}\vecu_1 R\vecu_2=R^{-1}\vecu_1 R\vecu_2 (PR)}
from (\ref{bra-com}) so that $f(PR)$ commutes past while the remaining $P$ in
$R'$ interchanges $1,2$. For example,
\eqn{act-covec-mat}{\pmatrix{x''&y''}=\pmatrix{x&y}\pmatrix{a'&b'\cr c'&d'}}
obeys the quantum plane relations if $x,y$ do and $a',b',c',d'$ are a copy of
the braided-matrices, and provided we remember the braid-statistics computed
from (\ref{bra-mat-covec-stat}). We proved the result in \cite{Ma:lin} for the
Hecke case where $R'\propto R$, but the general proof given here is almost the
same.

Likewise, remembering the braid-statistics the
tensor product of a braided-vector and a braided-covector is a braided-matrix,
at least in the Hecke case\cite{Ma:lin}. Thus
\eqn{proj-mat}{\vecv\vecx=\pmatrix{v^1x_1&\cdots &v^1x_n\cr \vdots&&\vdots\cr
v^nx_1&\cdots &v^nx_n}}
obeys the $B(R)$ relations (it is a rank-one braided matrix)\cite{Ma:lin}.
\note{ The proof is $R_{21}\vecv_1\vecx_1'
R\vecv_2\vecx_2'=R_{21}\vecv_1\vecv_2\vecx_1'\vecx_2'=\vecv_2\vecv_1
\vecx_1'\vecx_2'=\vecv_2\vecv_1\vecx_1'\vecx_2' R=\vecv_2\vecx_2'R_{21}
\vecv_1\vecx'_1 R$ as required. Here we used that $PR'\vecv_2\vecv_1=
\vecv_2\vecv_1$ and $\vecx_2\vecx_1 PR'=\vecx_2\vecx_1$ and note that
(at least formally) $R=Pf^{-1}(PR')$ for some function $f^{-1}$, so that
$R_{21}\vecv_1\vecv_2=f^{-1}(1)\vecv_2\vecv_1$. Similarly on the
$\vecx'_1\vecx'_2$ side in the reverse direction. This is essentially the
same as the Hecke case treated in \cite{Ma:lin}.}

This completes our summary of braided linear algebra in the most direct
language. In the following two chapters our goal is to introduce a slightly
more abstract way of working with and thinking about these braided objects. Our
first goal is to find a systematic way of obtaining these various
braid-statistics relations as in
(\ref{bra-mat-covec-stat})-(\ref{bra-mat-vec-stat}) etc. Since all braided
objects will enjoy some braid-statistics with all other braided objects, we
obviously need a general way of working these out. We also have to be sure that
everything is consistent (for example, one can show that the composite object
$\vecu''=\vecu\vecu'$ has the same braid-statistics as $\vecu$ or $\vecu'$ in
(\ref{bra-mat}) so that we can make higher products). We will formulate these
braid statistics by means of a braided-transposition operator $\Psi$ and
explain how this is obtained in a systematic way, often in terms of
$R$-matrices. This is our goal in the next chapter. Secondly, it is useful to
distinguish carefully between the original matrix algebra $B(R)$ and the copies
of it that are realised in terms of products of generators in $B(R)\und\tens
B(R)$ etc. Mathematically, this realization forms the {\em braided-coproduct
map} $\und\Delta: B(R)\to B(R)\und\tens B(R)$. It takes the matrix form
\eqn{B(R)-coprod}{\und\Delta u^i{}_j=u^i{}_k\tens u^k{}_j}
equivalent to the matrix multiplication in (\ref{bra-mat}). The underlines are
to remind us of the non-commuting braid statistics. Similar remarks apply for
the braided-vectors and braided-covectors. Thus the realization corresponding
to covector addition is the map $\und\Delta:V^*(R')\to V^*(R')\und\tens
V^*(R')$ given by
\eqn{V^*(R)-coprod}{\und\Delta x_i=x_i\tens 1+1\tens x_i.}
This leads us into a study of abstract braided-bialgebras and braided-Hopf
algebras in Chapter~4.

We see from this outline that our use of $R$-matrices in defining
braided-linear algebra really is quite different from their traditional use in
quantum inverse scattering (where they lead to quantum groups $A(R)$ etc).
Thus, it is not the case that all results concerning $R$-matrices arise in
inverse scattering. Quite simply, in QISM the QYBE leads to an associative
quantum-commutativity, while in our work the QYBE leads to an associative
braided-tensor product. Of course, the QYBE are well known to lead to braid
relations: our idea is to build these into the algebra  from the start in the
form of braid statistics.

\section{Braided Categories and the Unification of Covariance and Statistics}

In this chapter we formalise the notion of braid-statistics evident in the
examples of the braided-matrices above, by means of the notion of braided
tensor categories. Basically, this just means a collection of objects with a
braided-transposition law $\Psi$ obeying a number of consistency conditions. By
formalising it in this way, we keep track of those consistency conditions and
at the same time learn how to solve them. They need not come only from an
$R$-matrix as in the last chapter.

Let us recall that our goal is to generalise the notion of supersymmetry to the
braided case. In this chapter we concentrate on the first step, which is to
generalise the notion of vector-spaces and super-vector spaces. This is what a
braided category is from our point of view. Mathematically, it means that
everything is $\Z_2$-graded. On the other hand, in physics there is a slightly
different notion of covariance under a group. We shall see that when these
notions are generalised, which we do by means of quantum groups, they become
the same: the notion of quantum-group graded and quantum group-covariant
mathematically coincide. Hence we shall see that {\em the notion of
$Z_2$-grading (bose-fermi statistics) and the notion of group covariance are
unified by the notion of quantum symmetry}. This is an important unification in
physics made possible by quantum group technology. This point of view has been
developed in \cite[Sec. 6]{Ma:exa} and is the main physical lesson of the
present chapter.

Braided tensor categories have been formalised by category theorists in
\cite{JoyStr:bra} as well as being known in other contexts such as in the
theory of knot invariants and in connection with quantum groups. Firstly, a
category $\CC=\{V,W,Z,\cdots\}$ just means a collection of objects $V,W,Z$ etc
and a specification of what are the allowed morphisms $\phi:V\to Z$ etc between
them (in concrete situations they are maps between objects of some specified
form). One should be able to compose morphisms in an obvious way. So,
categories are nothing to be afraid of: they are just a specification of what
kind of objects and maps we intend to deal with. A braided tensor category is
$(\CC,\tens,\Phi,\Psi)$ where $\CC$ is a category, $\tens$ is a tensor product
between any two objects (and between any corresponding morphisms),
$\Phi_{V,W,Z}:V\tens (W\tens Z)\to (V\tens W)\tens Z$ is a collection of
isomorphisms expressing associativity of the tensor product between any three
objects, and $\Psi_{V,W}:V\tens W\to W\tens V$ (the braided transposition) is a
collection of isomorphisms expressing commutativity of the tensor product
between any two objects. In addition, there is a unit object $\und 1$ with
$V\tens\und 1\isom V\isom \und 1\tens V$. In our examples below, $\Phi$, the
unit object and its associated maps are typically the obvious ones, so we will
not write them too explicitly. However, they should be understood in all
formulae. By contrast, $\Psi$ will typically be non-trivial and so we emphasise
this.

These various collections of maps all fit together into a consistent framework.
The consistency conditions for $\Psi$  are of two types and modeled on the idea
that it behaves like usual transposition or super-transposition. Firstly,
$\Psi_{V,W}$ should be well-behaved under any morphisms of  $V$ or $W$ to any
other object,
\eqn{Psi-funct}{\Psi_{Z,W}(\phi\tens\id)=(\id\tens\phi)\Psi_{V,W}\
\forall\phi\matrix{\scriptstyle V\cr\downarrow\cr \scriptstyle Z},\qquad
\Psi_{V,Z}(\id\tens\phi)=(\phi\tens\id)\Psi_{V,W}\ \forall
\phi\matrix{\scriptstyle W\cr\downarrow\cr \scriptstyle Z}.}
One says that the collection is {\em functorial}. Secondly, it should be
well-behaved under tensor products of objects,
\eqn{Psi-hex}{\Psi_{V\tens W,Z}=\Psi_{V,Z}\Psi_{W,Z},\qquad \Psi_{V,W\tens
Z}=\Psi_{V,Z}\Psi_{V,W}.}
If we put in $\Phi$ and write these conditions as maps, they look like
hexagons. They are enough to imply also that $\Psi_{\und 1,V}=\id=\Psi_{V,\und
1}$ for the trivial object. $\Phi$  is also functorial and obeys a famous
pentagon condition.

These conditions (\ref{Psi-funct}-(\ref{Psi-hex}) are just the obvious
properties that we take for granted when transposing ordinary vector spaces or
super-vector spaces. In these cases $\Psi$ is the twist map $\Psi_{V,W}(v\tens
w)=w\tens v$ or the supertwist
\eqn{Psi-sup}{\Psi_{V,W}(v\tens w)=(-1)^{|v| |w|} w\tens v}
on homogeneous elements of degree $|v|,|w|$. The form of $\Psi$ in these
familiar cases does not depend directly on the spaces $V,W$ so we often forget
this. But in principle there is a different map $\Psi_{V,W}$ for each $V,W$ and
they all connect together as explained.

For example, the hexagons (\ref{Psi-hex}) say that if we transpose something in
$V\tens W$ past something in $Z$, we obtain the same result as first
transposing the part in $W$ with that in $Z$, and then the part in $V$.
Similarly on the other side. On the other hand, there is one crucial property
familiar for transpositions or super-transpositions which we do {\em not}
suppose. Usually, when transpositions or super-transpositions are applied
twice, they give the identity. Such a category is a {\em symmetric tensor
category}. By contrast, in our case we do not suppose that
$\Psi_{W,V}\Psi_{V,W}=\id$ and so must distinguish carefully between
$\Psi_{V,W}$ and $\Psi^{-1}_{W,V}$. They are both morphisms $V\tens W\to W\tens
V$. A convenient shorthand for doing this is to write them as morphisms
downwards and as braids rather than single arrows. Thus, we use the notation,
\eqn{Psi-bra}{\epsfbox{psi.eps}}
\noindent Thus, $\Psi$ behaves more like a braid crossing (and $\Psi^{-1}$ an
inverse braid crossing) than a
usual transposition (hence the name). It gives an action of the braid group on
tensor products of objects, rather than of the symmetric group. The coherence
theorem for braided categories says that if we are given any two composites of
the $\Psi, \Phi $ etc and write them as  braids according to (\ref{Psi-bra})
(suppressing the $\Phi$) then the compositions are the same if the
corresponding braids are topologically the same. This means that it does not
matter too much in what order we make a series of braided-transpositions -- the
result is the same if the corresponding braids are the same. Thus, the order or
history behind a series of braided-transpositions does matter to some extent
(unlike the situation with usual permutations) but only up to topology. This is
the novel feature in the braided case. Much of algebra involves permuting
objects past each other and this aspect of algebra is now replaced by topology.

This diagrammatic notation is a powerful one. In terms of it, we can write the
hexagon conditions (\ref{Psi-hex}) as
\eqn{Psi-hex-bra}{\epsfbox{hexagons.eps}}
\noindent where on the left of each equation we have extended our notation by
writing the strand for a composite object
such as $V\tens W$ or $W\tens Z$ as a pair of strands, one for each factor. The
content of (\ref{Psi-hex}) is that we can do this and and thereby include
(\ref{Psi-hex}) in our rule that topologically identical diagrams correspond to
the same resulting morphism.

Finally, we can write any other morphisms such as $\phi:V\to Z$ etc as nodes in
a strand connecting $V$ to $Z$. We write all morphisms downwards. A morphism to
or from tensor products will have multiple strands into or out of the node. In
these terms, the functoriality condition (\ref{Psi-funct}) comes out as
\eqn{Psi-funct-bra}{\epsfbox{functorial.eps}}
\noindent Thus functoriality means in our topological notation just that a node
(of any valency) can be pulled through a braid crossing (similarly for
$\Psi^{-1}$ for inverse braid crossings).

Thus, the main operations of tensor products and transpositions for vector
spaces and super-vector spaces are generalised to the braided case. Unlike
these cases, the order now matters to some extent and we have to be careful
that such braided-transpositions dont get tangled up. One other feature of
vector spaces and super-vector spaces that we will need is that of dual linear
spaces. Thus for every object $V$, there should be a dual $V^*$. In fact one
must distinguish carefully between left-duals and right-duals since they can
only be related via some further structure involving $\Psi$. We concentrate on
the left-duals (the right-handed ones are similar). Then the properties we need
are that for every object $V$ there should be another object $V^*$ and
morphisms $\ev_V:V^*\tens V\to \und 1$, $\coev_V:\und 1\to V\tens V^*$. For
vector spaces and super-vector spaces the trivial object $\und 1=\C$ and these
maps are given by
\eqn{ev-coev}{\ev(f\tens v)=f(v),\quad \coev(\lambda)=\lambda\sum_a e_a\tens
f^a}
where $v\in V,f\in V^*$, $\{e_a\}$ is a basis of $V$ and $\{f^a\}$ is a dual
basis. The abstract characteristic property of these maps is that
\eqn{rduals}{(\ev_{V}\tens\id)(\id\tens\coev_V)=\id_{V^*},\qquad
(\id\tens\ev_V)(\coev_V\tens\id)=\id_V.}
These are then enough to imply the most important of the familiar properties of
duals such as the ability to dualise morphisms. In the diagrammatic notation we
suppress $\und 1$ entirely so that the maps $\ev,\coev$ and the left-duality
condition (\ref{rduals}) appear as
\eqn{rduals-bra}{\epsfxsize=5.7in \epsfbox{rduals.eps}}
Thus the condition means in topological terms that an $S$-bend for $V^*$ and a
mirror-S-bend for $V$ can be straightened out by pulling. Some care is needed
however because the mirror images of these diagrams are not true (for them one
would need right-duals). A braided tensor category with both left and right
duals, suitably compatible, is often called a modular tensor category.

It is easy to see that these conditions hold for usual transpositions and
super-transpositions. Another source is to take for $\CC$ the category of
representations of a group. These have a commutative tensor product, with
$\Psi$ given by the usual vector-space transposition. On the other hand, a
fundamental property of strict quantum groups (with universal $R$-matrix) is
that they too can be used in this way. Thus, let us recall that a strict
quantum group means for us a quasitriangular Hopf algebra. This is
$(H,\Delta,\eps,S,\CR)$ where $H$ is an algebra, $\Delta:H\to H\tens
H$ the coproduct homomorphism, $\eps:H\to \C$ the counit, $S:H\to H$ the
antipode and $\CR$ the quasitriangular structure or `universal $R$-matrix'
obeying \cite{Dri}
\eqn{univR}{\nqquad (\id\tens\Delta)(\calR)=\calR_{13}\calR_{12},\
(\Deltaop\tens \id)(\calR)=\calR_{23}\calR_{13},\ \Deltaop=\calR(\Delta(\
))\CR^{-1}}
where $\calR_{12}=\calR\tens 1$ etc, and $\Deltaop$ is the opposite coproduct.
We have written the middle axiom in a slightly unconventional form but one that
generalises below. Here and throughout the paper $\C$ can be replaced by a
field of suitable characteristic or (with due care) a commutative ring.

Then a well-known theorem about strict quantum groups (and key to their
applications in knot theory) is that the category $\CC=\Rep(H)$ of
$H$-representations is a braided tensor one. The objects are vector spaces on
which $H$ acts, the morphisms are the $H$-interwiners, the associativity and
duals are the standard ones as for vector spaces. Here the tensor product
representation $V\tens W$ is given by the action of $\Delta(H)\subset H\tens
H$, the first factor acting on $V$ and the second factor on $W$. Finally, the
braiding is given by
\eqn{Psi-R}{\Psi_{V,W}(v\tens w)=P(\CR\la(v\tens w))}
where $\la$ is the action of $\CR\in H\tens H$ with its first factor on $V$ and
its second factor on $W$, and this is then followed by the usual vector-space
transposition $P$. The reason we need to act first by $\CR$ is that $P$ alone
would not be an intertwiner. Thus $h\la\Psi(v\tens w)=(\Delta h)\la P(\CR\la
(v\tens w))=P((\Deltaop h)\CR\la (v\tens w))=P(\CR(\Delta h)\la (v\tens
w))=\Psi(h\la(v\tens w))$ in virtue of the last of (\ref{univR}). It is easy to
see that the first two of (\ref{univR}) likewise just correspond to the
hexagons (\ref{Psi-hex}) or (\ref{Psi-hex-bra}). Functoriality is also easily
shown. For an early treatment of this topic see \cite[Sec. 7]{Ma:qua}. Note
that if $\CR_{21}=\CR^{-1}$ (the triangular rather than quasitriangular case)
we have $\Psi$ symmetric rather than braided.

\begin{propos}\cite[Sec. 6]{Ma:exa} Let $H=\Z_2'$ denote the quantum group
consisting of the group Hopf algebra of $\Z_2$ (with generator $g$ and $g^2=1$)
and a non-standard triangular structure,
\eqn{sup-R}{\Delta g=g\tens g,\quad \eps g=1,\quad Sg=g,\quad \CR=2^{-1}(1\tens
1+g\tens 1+1\tens g-g\tens g).}
Then $\CC=\Rep(\Z_2')={\rm SuperVec}$ the category of super-vector spaces.
\end{propos}
\proof One can easily check that this $\Z_2'$ is indeed a quasitriangular (in
fact, triangular) Hopf algebra. Hence we have a (symmetric) tensor category of
representations. Writing $p={1-g\over 2}$ we have $p^2=p$ hence any
representation $V$ splits into degree $V_0\oplus V_1$ according to the
eigenvalue of $p$. We can also write $\CR=1-2p\tens p$ and hence from
(\ref{Psi-R}) we compute $\Psi(v\tens w)=P(\CR\la (v\tens w))=(1-2p\tens
p)(w\tens v)=(1-2|v||w|)w\tens v=(-1)^{|v||w|}w\tens v$ as in (\ref{Psi-sup}).
\endproof

So this non-standard quantum group $\Z_2'$ (non-standard because of its
non-trivial $\CR$) recovers the category of super-spaces with its correct
symmetry $\Psi$. On the other hand, there are plenty of other quasitriangular
Hopf algebras $H$ we could use here. For example, in \cite{Ma:any} we have
introduced the non-standard quasitriangular Hopf algebra
$\Z_n'$ with generator $g$ and relation $g^n=1$ and
\eqn{any-R}{\Delta g=g\tens g,\quad \eps g=1,\quad S g=g^{-1},\quad
\CR=n^{-1}\sum_{a,b=0}^{n-1} e^{-{2\pi\imath ab\over n}}g^a\tens g^b.}
The category $\CC_n=\Rep(\Z_n')$ consists of vector spaces that split as
$V=\oplus_{a=0}^{n-1} V_a$ with the degree of an element defined by the action
$g\la v=e^{2\pi\imath |v|\over n}v$. From (\ref{Psi-R}) and (\ref{any-R}) we
find
\eqn{Psi-any}{\Psi_{V,W}(v\tens w)=e^{2\pi\imath\vert v\vert \vert w\vert\over
n}w\tens v.}
Thus we call $\CC_n$ the category of {\em anyonic vector spaces} of fractional
statistics ${1\over n}$, because just such a braiding is encountered in anyonic
physics. The case $n=2$ is that of superspaces. For $n>2$ the category is
strictly
braided in the sense that $\Psi\ne\Psi^{-1}$. There are natural anyonic traces
and anyonic dimensions generalizing the super-case\cite{Ma:any}
\eqn{any-dim}{\und\dim V=\sum_{a=0}^{n-1} e^{-{2\pi\imath a^2\over n}}\dim
V_a,\qquad \und{\rm Tr}(f)=\sum_{a=0}^{n-1}e^{-{2\pi\imath a^2\over n}} \trace
f|_{V_a}.}
We see that the category is generated by the quantum group $\Z_n'$.

Note that these quantum groups are discrete and nothing whatever to do with
usual $q$-deformations. Quite
simply, this use of the mathematical structure of quasitriangular Hopf algebras
as generating statistics is
different from how they arose in QISM. On the other hand, there is nothing
stopping us going to the other
extreme and taking $H=U_q(g)$. For example, if $H=U_q(sl_2)$ the role of $\Z_2$
or $\Z_n$-grading is now played by
the spectral decomposition into irreducibles. Thus $V=\oplus n_i V_i$ where the
$V_i$ are the spin $i=0,\h,1$ etc.  representations. The braiding $\Psi$ from
(\ref{Psi-R}) is now given by the direct sum of the corresponding $R$-matrices
for each spin, which in turn can be reduced to products of the fundamental
$SL_q(2)$ $R$-matrix.

We call the quantum group $H$ used in this way the {\em statistics generating
quantum group}\cite[Sec. 6]{Ma:qua}. Each generates a braided category or
`universe' within which we can work. By this reasoning we can immediately
generalise the notion of super-algebras. Thus, a {\em braided-algebra} means an
algebra $B$ living in a braided tensor category. We mean by this that there is
an object $B$ in the category and the product and unit maps
\eqn{bra-alg}{\cdot:B\tens B\to B,\qquad \eta:\und 1\to B}
are morphisms in the category.

Now we switch to another topic, that of covariance under a group. Recall that
an algebra $B$ is $G$-covariant if the group $G$ acts on $B$ and $g\la
(bc)=(g\la b)(g\la c)$, $g\la 1=1$ for all $g\in G$ and $b,c\in B$. Here $\la$
denotes the action. The natural generalization of this to any Hopf algebra $H$
is that of an $H$-covariant algebra (or $H$-module algebra). This means an
algebra $B$ on which $H$ acts according to
\eqn{H-modalg}{ h\la (b\cdot c)=\cdot((\Delta h)\la(b\tens c)),\quad h\la
1=\eps(h)1}
where $\Delta h$ in $H\tens H$ acts with the first factor on $b$ and the second
factor on $c$, and we have emphasised the product $\cdot$ in $B$. This notion
also includes the notion of $g$-covariant where $g$ is a Lie algebra. There
$\Delta\xi=\xi\tens 1+1\tens\xi$ for $\xi\in g$ and so the covariance condition
(\ref{H-modalg}) becomes $\xi\la (bc)=(\xi\la b)c+b(\xi\la c)$ as usual (groups
and Lie algebras are unified when we work with Hopf algebras).

\begin{lemma} Let $(H,\CR)$ be a quasitriangular Hopf algebra and
$\CC=\Rep(H)$. An algebra $B$ lives in $\CC$ (is a braided algebra) iff it is
$H$-covariant in the sense of (\ref{H-modalg}).
\end{lemma}
\proof That $B$ lives in the category as in (\ref{bra-alg}) means in the
present setting that the product and unit maps are intertwiners for the action
of $H$ (they are morphisms in $\Rep(H)$). This means $h\la (b\cdot
c)=\cdot(h\la (b\tens c))$. But the action of $H$ on $B\tens B$ is in the
tensor product representation, so $h\la (b\tens c)=(\Delta h)\la (b\tens c)$.
\endproof

This lemma in conjunction with the above analysis of super-spaces etc expresses
the unification of two concepts made possible by quantum group theory. Thus
\eqn{unif}{ B\in \Rep(H)\quad{\rm means}\quad  \cases{{\rm
super-algebra}&$\quad H=\Z_2'$\cr {\rm G-covariant\  algebra}&$\quad H=\C G$}}
Moreover, there are plenty of other strict quantum groups (quasitriangular Hopf
algebras) that one may take here, ranging from the anyonic
statistics-generating Hopf algebra $\Z_n'$ to the more standard $U_q(g)$. Each
can be interpreted either way, as covariance (a quantum symmetry) or as
generating statistics.

\section{Diagrammatic Methods for Braided-Hopf Algebras}

The unification of statistics and covariance in the last chapter means that all
results about
braided algebra, developed by analogy with super-symmetry, can then be
reinterpreted in terms of
results about quantum-group covariant systems. We will make such applications
in later chapters. Our goal
in the present chapter is to develop this braided-algebra in analogy with the
theory of
super-algebras. This provides the mathematical underpinning of the examples
such as the
braided-matrices in Chapter~2, as well as providing some powerful diagrammatic
tools for working
with such objects.

Just as two super algebras $B,C$ have a super-tensor-product superalgebra,
$B\und\tens C$, containing
$B,C$ as mutually super-commuting sub-superalgebras, so we have now the
fundamental lemma:
\begin{lemma}  Let $B,C$ be two algebras living in a braided category (two
braided algebras as in (\ref{bra-alg})). There is a {\em braided tensor product
algebra} $B\und\tens C$, also living in the braided category. It is built on
the
object $B\tens C$ but with the product law
\eqn{tensprod}{(a\tens c)(b\tens d)=a\Psi_{C,B}(c\tens b)d\quad\forall a,b\in
B,\ c,d\in C\quad \epsfbox{tensprod.eps}}
where the first is a concrete description and the second an abstract one in
terms of morphisms.
\end{lemma}
\proof The best proof that this product law is associative is the diagrammatic
one in \cite{Ma:sta}
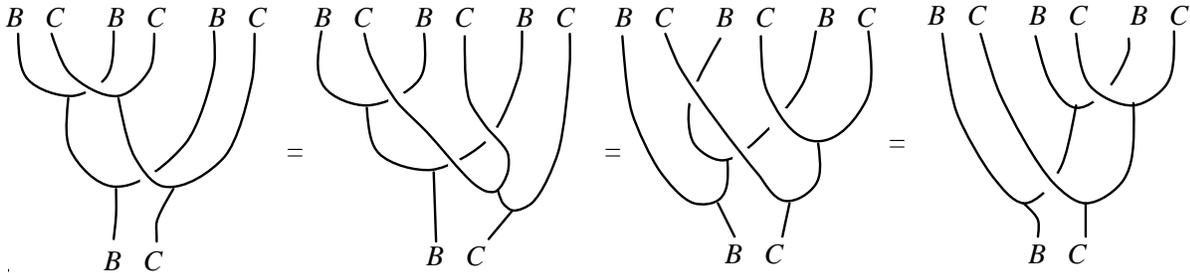
\[ \epsfbox{prod-assoc.eps}\]
The first step uses functoriality as in (\ref{Psi-funct-bra}) to pull the
product morphism through the braid crossing. The second equality uses
associativity of the products in $B,C$ and the third equality uses
functoriality again in reverse. The product is manifestly a morphism in the
category (is covariant) because it is built out of morphisms (covariant maps).
Finally, the unit is the tensor product one because the braiding is trivial on
$\und 1$. \endproof

 In the concrete case the braided tensor product is generated by $B=B\tens 1$
and $C=1\tens C$ and an exchange law between the two factors given by $\Psi$.
This is because $(b\tens 1)(1\tens c)=(b\tens c)$ while $(1\tens c)(b\tens
1)=\Psi(c\tens b)$. Another notation is to label the elements of the second
copy in the braided tensor product by ${}'$. Thus $b\equiv (b\tens 1)$ and
$c'\equiv (1\tens c)$. Then if $\Psi(c\tens b)=\sum b_k \tens c_k$ say, we have
the braided-tensor product  relations
\eqn{btens-stat}{ c'b\equiv (1\tens c)(b\tens 1)=\Psi(c\tens b)=\sum b_k\tens
c_k\equiv \sum b_k c'_k}
which is the notation used in Chapter~2. This also makes clear why we call
$\Psi$ the braid-statistics and why the lemma generalizes the notion of
super-tensor product. From the unification of statistics and covariance in
Lemma~3.2, we have equally well,

\begin{corol} Let $H$ be a strict quantum group. Given two $H$-covariant
algebras $B,C$, we have another $H$-covariant algebra $B\und\tens C$.
\end{corol}

 We note that there is an equally good {\em opposite braided tensor product}
with the inverse braid crossing in Lemma~4.1. For any braided category $\CC$
there is another mirror-reversed braided category $\bar{\CC}$ with
\eqn{psi-barC}{\bar{\Psi}_{V,W}=\Psi_{W,V}^{-1}}
in place of $\Psi_{V,W}$ (braids and inverse braided interchanged), and the
opposite braided tensor product algebra is simply the braided tensor product
algebra in $\und\CC$.

This lemma and its corollary have many applications, as we shall see in
Chapter~6. It is also the fundamental lemma for us now because it enables us to
define the notion of braided group as a generalization of supergroups and
super-enveloping algebras. Let us recall that in place of working with groups
and Lie algebras we can work equivalently with the group Hopf algebras and
cocommutative enveloping Hopf algebras that they generate (cocommutative means
that $\Deltaop=\Delta$). The same holds in the super-case where one can work
with super-cocommutative enveloping super-Hopf algebras. Thus, we will
introduce braided groups formally as (in a certain sense braided-cocommutative)
braided-Hopf algebras.

Recall that the key feature of a Hopf algebra is that it has a realization in
its own tensor products. In the braided case we require a realization in its
own braided tensor products. Thus, a braided-Hopf algebra (a Hopf algebra
living in a braided category $\CC$) is $(B,\Delta,\eps,S)$ where $B$ is a
braided algebra as in (\ref{bra-alg}) and $\Delta:B\to B\und\tens B$,
$\eps:B\to\und 1$ are algebra homomorphisms where $B\und\tens B$ has the
braided tensor product algebra structure. In addition $\Delta,\eps$ are
coassociative as usual, and $S:B\to B$ obeys the usual axioms of an antipode.
In diagrammatic form, the axioms are
\eqn{hopf-ax}{\epsfbox{hopf-ax.eps}}
If there is no antipode then we speak of a braided-bialgebra or bialgebra in a
braided category.

This gives us the notion of a group-like or Hopf algebra-like object in a
braided category. In what follows we shall see that these braided objects
really behave just like usual groups or Hopf algebras (or their
super-versions). For example, recall that the antipode behaves like a group
inverse (for a group algebra $Sg=g^{-1}$) with the result of course that $S$ is
an anti-algebra homomorphism.

\begin{lemma} For a braided-Hopf algebra $B$, the braided-antipode obeys
$S(b\cdot c)=\cdot\Psi(Sb\tens Sc)$ and $S(1)=1$, or more abstractly,
$S\circ\cdot=\cdot\circ\Psi_{B,B}\circ(S\tens S)$ and $S\circ\eta=\eta$. Also,
$\Delta\circ S=(S\tens S)\circ\Psi_{B,B}\circ\Delta$ and $\eps\circ S=\eps$.
\end{lemma}
\proof In diagrammatic form the proof is\cite{Ma:tra}
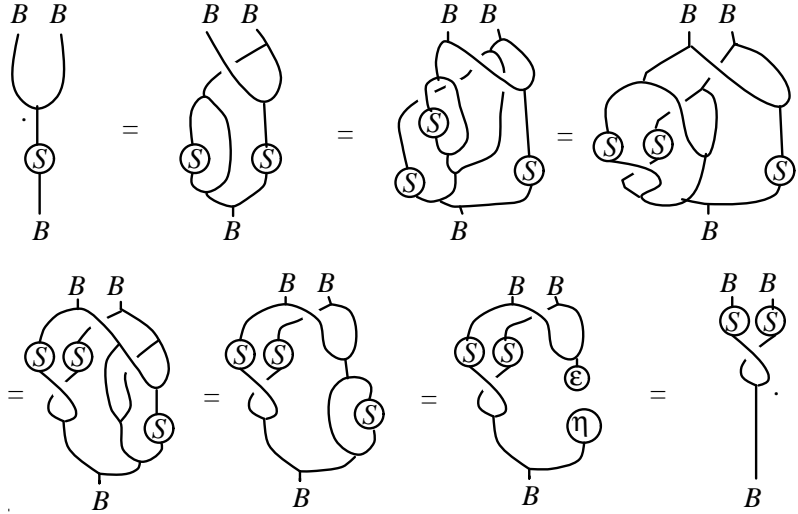
\[\epsfbox{ant-ant.eps}\]
In the first two equalities we have grafted on some circles containing the
antipode, knowing they are trivial from (\ref{hopf-ax}). We then use the
coherence theorem to lift the second $S$ over to the left, and associativity
and coassociativity to reorganise the branches. The fifth equality uses the
axioms (\ref{hopf-ax}) for $\Delta$. For the second part of the lemma, turn the
diagram-proof upside-down and read it again.
\endproof

To see that this is a useful notion, recall that any group acts on itself by
the adjoint action $\Ad_h(g)=hgh^{-1}$. Remembering that $\Delta h=h\tens h$
for a group Hopf algebra, the steps here are to split $h$ using the coproduct
$\Delta$, apply $S$ to the second factor and multiply up with $g$ in the
middle. Writing this as maps in our braided case we have:

\begin{propos} Every braided-Hopf algebra $B$ acts in itself by a
braided-adjoint action $Ad=(\cdot\tens\cdot)(\id\tens\Psi_{B,B})(\id\tens
S\tens\id)(\Delta\tens\id)$.
\end{propos}
\proof The diagrammatic representation and proof is
\eqn{Ad}{\epsfbox{Ad.eps}\quad\qquad\epsfbox{Ad-act.eps}}
Here the first step is functoriality while second equality uses the result in
the preceding lemma. We then use the bialgebra axiom for $\Delta$ in
(\ref{hopf-ax}). We have adopted the notation of combining repeated products
into a single node, knowing (from associativity) that only the order of inputs
into the combined product matters, not its original tree structure.
\endproof

We can continue to prove familiar properties for the braided-adjoint action.
For example, in the group case it respects its own algebra structure. The
corresponding condition for an algebra $C$ in a braided-category to be
braided-covariant under $B$ (a braided $B$-module algebra) is
\eqn{C-modalg}{\epsfbox{C-modalg.eps}\qquad B{\rm -Module\  Algebra}}
This is the braided analogue of the group covariance $g\la(bc)=(g\la b)(g\la
c)$ or the quantum covariance as in (\ref{H-modalg}). For the record, we can
also ask that a coalgebra $(C,\Delta,\eps)$ in the category is acted upon by
$B$. The corresponding condition is
\eqn{C-modcoalg}{\epsfbox{C-modcoalg.eps}\qquad B{\rm -Module\ Coalgebra}}

\begin{propos} The braided-adjoint action of $B$ on itself given in
Proposition~4.4 respects its product in the sense of (\ref{C-modalg}).
\end{propos}
\proof
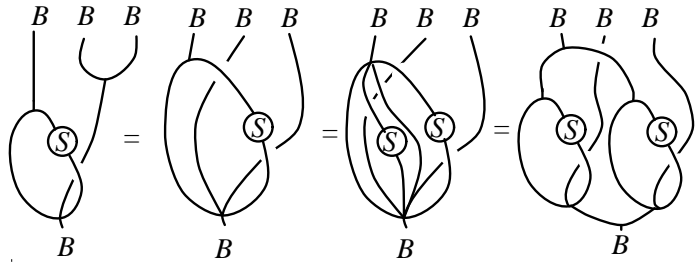
\[ \epsfbox{Ad-modalg.eps}\]
Here we used functoriality in the first step and grafted on a trivial circle
involving $S$ in the second step. We then used associativity and
coassociativity to organise the result. We see that (\ref{C-modalg}) is obeyed
with $\alpha={\rm Ad}$.
\endproof

Next, we come to the question of what we mean by a cocommutative (cf
commutative) Hopf algebra in a braided category.
The most naive idea would be to define $\Deltaop=\Psi_{B,B}\circ\Delta$ or
$\Deltaop=\Psi^{-1}_{B,B}\circ\Delta$ as an opposite Hopf algebra, and say that
$B$ is cocommutative if $\Deltaop=\Delta$. Unfortunately, in the truly braided
case the first does not give a Hopf algebra at all (things get too tangled up)
while the second gives

\begin{lemma} $B^{\rm cop}$ defined as the same algebra $B$ but with
braided-coproduct $\Deltaop=\Psi^{-1}_{B,B}\circ\Delta$  defines a bialgebra in
$\bar{\CC}$ ($\CC$ with the opposite braiding). It is a Hopf algebra in
$\bar{\CC}$ with  antipode $S^{\rm op}=S^{-1}$ if $S$ is invertible.
\end{lemma}
\proof We use the definitions and the axioms (\ref{hopf-ax}) for $B$,
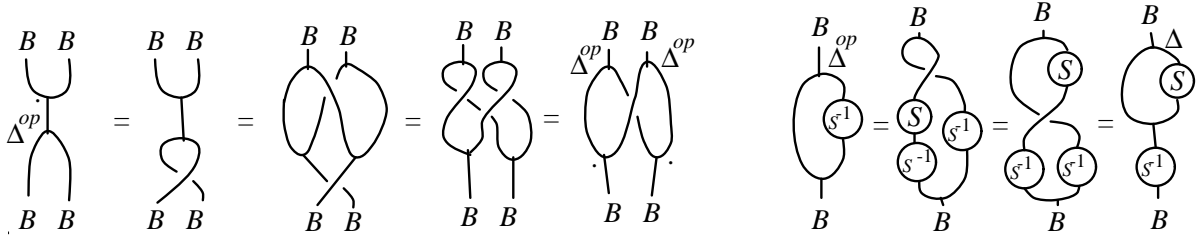
\[ \epsfbox{Bop.eps}\]
\endproof

Because of this, there is no intrinsic notion of opposite coproduct lying in
our original category, hence no intrinsic notion of cocommutativity in the
braided case. Our way out in the theory of braided groups is to realise that we
dont usually need cocommutativity in an intrinsic sense but only when the Hopf
algebra acts on things, i.e. relative to representations. The notion of a
braided-representation or braided-module $V$ is just the obvious one ($V$ is an
object in the category and $B$ acts on it by a morphism $B\tens V\to V$). Thus,
we can turn things around and (for any braided-Hopf algebra $B$) we say that
{\em $B$ behaves braided-cocommutatively with respect to a
braided-representation $V$} if
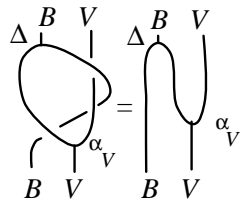
\eqn{V-cocom}{\epsfbox{V-cocom.eps}\qquad {\rm Braided\ Cocommutativity}}
In the symmetric (unbraided) case this reduces to
\eqn{V-cocom-tri}{\epsfbox{V-cocom-tri.eps}}
but in general we cannot untangle $V$ from $B$ and must work with the weak
notion.

\begin{lemma} If $B$ is braided-cocommutative with respect to the
braided-adjoint action (for example, this is true for the examples in
Chapter~5) then the braided-adjoint action respects the coproduct of $B$ in the
sense of (\ref{C-modcoalg}).
\end{lemma}
\proof
\[ \epsfbox{Ad-modcoalg.eps}\]
Here the first and second equalities both use the bialgebra axiom for $\Delta$
in (\ref{hopf-ax}). The third uses Lemma~4.3. The fourth uses functoriality to
drag the second $S$ up and over to the left, so that we can recognise that it
forms ${\rm Ad}$ in the fifth. The seventh uses the braided-cocommutativity
assumption (\ref{V-cocom}) for ${\rm Ad}$.
\endproof

This is a result usually reserved for honest groups and supergroups (or
cocommutative (super)-Hopf algebras) and its failure for general Hopf algebras
severely limits the use of the quantum-adjoint action. Here we see that the
requirement of cocommutativity with respect to $\Ad$ plays an analogous role.
This kind of cocommutativity then tells us that we have a braided group rather
than merely a braided-Hopf algebra. Formally then, we define a braided group as
$(B,\CO)$ where $B$ is a braided-Hopf algebra and $\CO$ is a useful class of
$B$-modules with respect to which $B$ behaves cocommutatively.

Next, let is recall that for any group or Hopf algebra we can tensor product
representations, and also dualise them using the group inverse. The same is
true in the braided case.

\begin{propos} Let $B$ be a braided-Hopf algebra. The braided tensor product of
two $B$-modules $V,W$ is given by the action of $B$ on $V\tens W$ via the
action of $\Delta(B)$ and remembering the braid-statistics $\Psi$. In concrete
terms this is
\[ b\la (v\tens w)=\sum b\Bo\la \Psi(b\Bt\tens v)\la w\]
where $\la$ denotes the relevant actions and $\Delta b=\sum b\Bo\tens b\Bt$ is
an explicit notation in the concrete case.
As morphisms this is  $\alpha_{V\tens
W}=(\alpha_V\tens\alpha_W)\Psi_{B,V}(\Delta\tens\id\tens\id)$.
\end{propos}
\proof
\eqn{tens-act}{\epsfbox{tens-act.eps}\qquad\quad
 \epsfbox{tensVW.eps}}
\endproof

\begin{lemma} If $V$ is a left braided $B$-module, then its left dual $V^*$ is
a right braided $B$-module by dualising.
\end{lemma}
\proof The dualization uses the $\ev$ map in (\ref{rduals}). In diagrammatic
form the resulting action $\alpha^*$ on $V^*$ is
\eqn{V*-mod}{\epsfbox{Vp-mod.eps}}
\endproof
\begin{lemma} If $V$ is a right $B$-module then it is also a left $B$-module
via the antipode $S$.
\end{lemma}
\proof We use the antipode to convert a right action to a left action (recall
from Lemma~4.3 that $S$ is some kind of anti algebra homomorphism). If
$\alpha^{\rm R}$ is our initial right action on $V$, the required left action
is
\eqn{V-rlmod}{\epsfbox{V-rlmod.eps}}
\endproof

The Proposition~4.8 says that we can tensor product braided-representations,
while using the lemmas, we see that any braided-representation $(V,\alpha_V)$
of a braided-Hopf algebra (a left $B$-module) has a contragradient or dual one
$(V^*,\alpha_{V^*})$ where $\alpha_{V^*}$ is given by feeding the result of
Lemma~4.9 into Lemma~4.10. This is just as for usual groups or Hopf algebras
where the inverse or antipode allows one to construct contragradient
representations.

On the other hand, a feature of true groups (as against general Hopf algebras)
is that the tensor product of their representations is symmetric under the
usual transposition of the underlying vector space. In the braided case we have
the analogous result with $\Psi$ in the role of the usual transposition. Thus,
if $B$ is braided-cocommutative with respect to $V$ then
\eqn{comtens}{\epsfbox{comtens.eps}\qquad {\rm Braided-Commutativity\ of\
Representations}}
from (\ref{V-cocom}) (by adding an action on $W$ to both sides in
(\ref{V-cocom})) and  is largely equivalent to it.  This is how (\ref{V-cocom})
was derived in \cite{Ma:bra}. These results thus demonstrate that our weak
braided-cocommutativity condition (\ref{V-cocom}) is an appropriate and useful
one.

We now turn to another topic, namely that of dual braided-Hopf algebras. Recall
that a novel feature of Hopf algebras is that the dual of a Hopf algebra is
also a Hopf algebra. For example, the dual of a group Hopf algebra is the
commutative algebra of functions on a group. Similarly in the super case. When
we work with matrix super-groups we are making use of this duality by
describing the super-commutative Hopf algebra of `functions' dual to the
corresponding enveloping super-Hopf algebra itself.

\begin{propos} If $B$ is a braided-Hopf algebra, then its left-dual $B^*$ is
also a braided-Hopf algebra with product, coproduct, antipode, counit and unit
given by
\eqn{B-dual}{\epsfbox{B-dual.eps}}
\end{propos}
\proof Associativity and coassociativity follow at once from coassociativity
and associativity of $B$. Their crucial compatibility property comes out as
\[ \epsfbox{dual-bialg.eps}\]
The antipode property comes out just as easily.
\endproof

Clearly, one can go on and develop all the usual theory of Hopf algebras in
this braided-diagrammatic way. We will content ourselves here with another
canonical construction, namely the braided version of the coregular action of a
Hopf algebra on its dual. This is needed in Chapter~7 in order to define
braided-vector fields (just as a group or Lie algebra acts on its own function
algebra by derivations).

The first step is to recall that an action of a Hopf algebra corresponds to a
coaction of its dual. A coaction is like an action but with the arrows
reversed. Thus, a left $B$-comodule is a map $\beta_V:V\to B\tens V$.

\begin{propos} If $V$ is a left $B$-comodule then it becomes a left
$B^*$-module by dualising.
\end{propos}
\proof Here we use $\ev$ in (\ref{rduals}). The diagrammatic form and proof is
\eqn{V-B*mod}{\epsfbox{V-Bpmod.eps}}
\endproof

Next, if $B$ coacts on some other algebra we can ask that $\beta_C:C\to
B\und\tens C$ is an algebra homomorphism to (a realisation in) the braided
tensor product as in Lemma~4.1. This is a {\em braided-comodule algebra},
\eqn{C-comodalg}{\epsfbox{C-comodalg.eps}\qquad B{\rm -Comodule\ Algebra}}
For example, the coproduct $\Delta:B\to B\tens B$ makes $B$ automatically a
left (and also a right) comodule algebra under itself. For the record, we can
also ask that a coalgebra $(C,\Delta,\eps)$ is preserved under a coaction of
$B$. The corresponding condition is
\eqn{C-comodcoalg}{\epsfbox{C-comodcoalg.eps}\qquad B{\rm -Comodule\
Coalgebra}.}

\begin{lemma} Let $C$ be a braided left $B$-comodule algebra as in
(\ref{C-comodalg}). Then $C$ is also a left $B^*{}^{\rm cop}$-module algebra in
$\bar{\CC}$ by dualising.
\end{lemma}
\proof We have seen that $B^*$ is a braided Hopf algebra in $\CC$ and hence by
Lemma~4.6 applied to $B^*$, we know that $B^*{}^{\rm cop}$ is a Hopf algebra in
the category $\bar{\CC}$ with reversed braiding. Also, we know from
Proposition~4.12 that a coaction $\beta$ of $B$ gives an action $\beta^*$ of
$B^*$. We compute
\eqn{C-B*modalg}{\epsfbox{C-Bpmodalg.eps}}
The first equality is the definition of $\beta^*$ and the second uses our
assumption (\ref{C-comodalg}) for $\beta$. We then use the definition of the
coproduct for $B^*$ in (\ref{B-dual}) and  recognise $\beta^*$ again. The right
hand side is just like the right-hand side of (\ref{C-comodalg}) except that we
have the opposite coproduct of $B^*$ and we have an inverse braid crossing as
$B^*$ passes $C$, rather than the crossing in (\ref{C-modalg}). Hence we have a
$B^*{}^{\rm cop}$-module algebra living in the opposite category $\bar{\CC}$.
\endproof

\begin{lemma} Let $C$ be a braided left $B^{\rm cop}$-module algebra in
$\bar{\CC}$. Then $C$ is also a right $B$-module algebra in $\CC$ by the use of
$S$.
\end{lemma}
\proof We convert the left $B$-module $\alpha$ to a right one $\alpha_R$ in a
way similar to Lemma~4.10. The proof is also similar, namely take the diagram
proof there, reflect it in a mirror about a vertical axis and put the braid
crossings that are reversed by this process back to their original form. We
show that the resulting right action $\alpha_R$ defines a right module algebra
(acts covariantly on the algebra $C$ from the right). We have
\eqn{C-lrmodalg}{\epsfbox{C-lrmodalg.eps}}
The first equality is functoriality while the second is our assumption that $C$
is a $B^{\rm cop}$-module algebra in $\bar{\CC}$ in the sense explained in the
preceding lemma (like (\ref{C-modalg}) but with $\Delta^{\rm op}$ rather than
$\Delta$ and a reversed braid crossing when $B$ passes $C$). We then use
Lemma~4.3 and recognise the result. The result is that $\alpha_R$ acts on $C$
according to a right-handed version of (\ref{C-modalg}). This right handed
version consists of (\ref{C-modalg}) reflected in a mirror about a vertical
axis and with its reversed braid crossing then restored.
\endproof

For example, if we apply Lemma~4.13 to the left-regular coaction of $B$ in
itself given by $\Delta$, we see that $B^*{}^{\rm cop}$ acts on $B$ from the
left, but in the opposite category. Also, feeding the result of this into
Lemma~4.14 applied to $B^*{}^{\rm cop}$, we see that every braided-Hopf algebra
$B$ becomes a right $B^*$-module algebra in our original category by
\eqn{right-vect}{ \epsfbox{right-vect.eps}}

In this chapter we have defined only the elementary theory of braided groups.
Other results (in our diagrammatic form) appeared in \cite{Ma:bos} and
\cite{Ma:tra} and we recall them now without proof. In the first we show that
if $C$ is a left $B$-module algebra as in (\ref{C-modalg}) then there is a {\em
cross product braided-algebra} $C\cocross B$. It is built on $C\tens B$ with
product given by the top left of
\eqn{C-cross}{ \epsfbox{C-cross.eps}}
We proved associativity etc. As an application we showed that if the $B$-module
algebra $C$ is also a braided-Hopf algebra and $B$ is cocommutative with
respect to $C$ then $C\cocross B$ is a braided-Hopf algebra with the
braided-tensor-coproduct. Again, this is a result usually formulated only for
group actions or actions of cocommutative Hopf algebras. It leads to the
bosonization result in Chapter~5.

Reflection of this left semidirect or cross product construction in the
vertical axis takes us into an analogous construction for a right $B$-module
algebra living in $\bar{\CC}$, and restoration of the reversed braid-crossings
gives us the analogous construction for a right $B$-module algebra in our
original category (such as obtained in the last lemma), as displayed at the top
right in (\ref{C-cross}).
For example, we have by this construction the {\em braided Weyl algebra}
$B^*\cross B$ using (\ref{right-vect}). This is the semidirect product of $B$
regarded as `position observables' by $B^*$ acting from the right as `braided
vector-fields'.

As well as the reflection principle about a vertical axis, we also encountered
above a reflection principle about a horizontal axis. This converts the theory
with left modules into one with left comodules living in the opposite category.
Restoring the reversed braid crossings gives us the theory for left comodules
in our original category. Since this applies to all the axioms, it applies to
all our results above obtained by diagram-proofs. For example, we have a
semidirect product coalgebra for a coalgebra $C$ crossed by a left coaction
obeying (\ref{C-comodcoalg}), displayed bottom left in (\ref{C-cross}).
Finally, by making both reflections (turning the page up-side-down) we have the
theory for right-comodules in our original category. Again, it is not necessary
to repeat all the diagram proofs: All the  results above about left modules
hold for right comodules (and vice versa) by turning the page upside down and
interchanging products and coproducts, evaluations and coevaluations etc. The
duality and Hopf algebra axioms (\ref{rduals}) and (\ref{hopf-ax}) are
unchanged by this. The notion of braided-cocommutativity with respect to a
module in (\ref{V-cocom}) now becomes the notion of {\em braided-commutativity}
with respect to a comodule. The left adjoint action becomes the right adjoint
coaction etc. This is the form of the theory actually used in Chapter~2. See
\cite{Ma:lin} for the details of the derivation.

Finally, in \cite{Ma:tra}, we study quantum-braided-groups, by which we mean a
braided-Hopf algebra equipped with a braided-universal $R$-matrix. The latter
is a morphism $\CR:\und 1\to B\tens B$ obeying
\eqn{B-univR}{\epsfbox{B-univR.eps}}
where $\Deltaop$ is any second Hopf algebra structure on $B$. Again, its
usefulness is limited to $B$-modules $V$ obeying the condition (\ref{V-cocom})
with $\Deltaop$ in place of the $\Delta$ on the left (representations in which
$\Deltaop$ behaves like an opposite coproduct). We show, for example, that the
class $\CO(B,\Deltaop)$ of all such $B$-modules is closed under tensor products
and itself a braided category in the quantum-braided-group case. If our initial
braiding is equivalent to some kind of quantization\cite{Ma:csta}, this is
equivalent to second-quantization. Again, there is a dual version of the theory
with comodules, which is the original setting in \cite{Ma:bg}.

\section{Transmutation and Bosonization}

In the above we have given a full account of the elementary part of the theory
of braided groups and braided matrices. In what remains we shall content
ourselves with the briefest of outlines, without proof, of some of the more
advanced theorems  and their applications. In the present chapter we shall
state the main results about transmutation and bosonization, and something of
the general principles involved. The full details are in
\cite{Ma:eul}\cite{Ma:bg}\cite{Ma:tra}\cite{Ma:bos}. In the next two chapters
we will conclude with a few illuminating examples and applications of interest
in physics. Readers who have had enough of abstract mathematics should proceed
directly to these chapters and skip the present one.

Recall that in quantum theory, and also in our algebraic work in Chapter~2,
there are really two sources of non-commutativity. The first is
non-commutativity of the algebra of observables resulting from quantization and
expressed in algebraic terms in the language of non-commutative geometry (or as
non-cocommutativity in the case of quantum enveloping algebras), while the
second is statistical non-commutativity expressed as non-commutation relations
between independent copies of the algebra. Thus super-groups and braided-groups
(of function algebra type) are viewed not as quantum objects (they are super-
or braided-commutative) but rather, the non-commutativity is of this second
type and expressed in Chapter~2 as braid statistics. Thus the two kinds of
non-commutativity are conceptually quite different.

The idea behind this chapter is that sometimes one can systematically trade one
of these kinds of non-commutativity for the other. Thus, we can exchange some
of the quantum-non-commutativity for statistical non-commutativity, and
vice-versa. This means that how we view an object, as a quantum group,
super-group or braided-group is to some extent a matter of choice. The category
we work in is a kind of `co-ordinate system' and one system may be better for
some purposes than another.

The precise formulation is as follows. Let $H_1{\buildrel f\over\to} H$ be a
pair of quantum groups with a Hopf algebra map between them (for example,
$H_1\subseteq H$). At least $H_1$ should have a universal R-matrix. As
explained in Chapter~3, it generates a braided category $\Rep(H_1)$.

\begin{theorem}\cite{Ma:bra}\cite{Ma:tra} $H$ can be viewed equivalently as a
braided-Hopf algebra $B(H_1,H)$ living in the braided category $\Rep(H_1)$ (so
with braid statistics induced by $H_1$). Here
\eqn{B(H)}{ B(H_1,H)=\cases{H&{\rm as\ an\ algebra}\cr
 \und\Delta,\ \und S &{\rm modified\ coproduct\ and\ antipode}}}
If $H$ has a universal R-matrix then $B(H_1,H)$ has a braided-universal
$R$-matrix as in  (\ref{B-univR}), given by the ratio of the universal
R-matrices of $H,H_1$.
\end{theorem}

The explicit formulae are as follows. Firstly, the action of $H_1$ is the
induced quantum adjoint action $h\la b=Ad_{f(h)}(b)=\sum f(h\o) b f(Sh\t)$
where $\Delta h=\sum h\o\tens h\t$ is an explicit standard notation for the
coproduct. Using this, the braided coproduct, braided-antipode and braided
quasitriangular structure (or braided-universal R-matrix) are
\eqn{B(H)-hopf}{ \und\Delta b=\sum b\o f(S\CR_1\ut)\tens \CR_1\uo\la b\t,\quad
\und S b=\sum f(\CR_1\ut)S( \CR_1\uo\la b)}
\eqn{B(H)-univR}{ \und\CR=\sum \rho\uo f(S\CR_1\ut)\tens \CR_1\uo\la\rho\ut}
where $\rho= f(\CR_1^{-1})\CR$ is the ratio as promised and $\uo,\ut$ refer to
the component factors in a tensor product. There is also a specific opposite
coproduct characterised by
\eqn{B(H)-op}{\sum \Psi(b_{\und {(1)_{\rm op}}}\tens Q_1\uo\la
b_{\und {(2)_{\rm op}}})f(Q_1\ut)=\sum b_{\und {(1)}}\tens b_{\und {(2)}}}
where $Q_1=(\CR_1)_{21}(\CR_1)_{12}$ and $f(Q_1\ut)$ multiplies the result of
$\Psi$ from the right as $(1\tens f(Q_1\ut))$. The underlines in the
superscripts
are to remind us
that we intend here the braided-coproducts $\und\Delta$ and $\und\Deltaop$. The
equation  can also be inverted to give an explicit formula for $\und\Deltaop$.
That these formulae obey the axioms (\ref{hopf-ax}) and (\ref{B-univR}) of
Chapter~4 is verified explicitly in \cite{Ma:tra}.

\begin{corol}\cite{Ma:any} Let $H$ be a quantum group containing a group-like
element $g$ of order $n$ (so $g^n=1$ and $\Delta g=g\tens g$). Then $H$ has a
corresponding anyonic version $B$. It has the same algebra and
\eqn{trans-any-hopf}{\und\Delta b=\sum b\o g^{-\vert b\t\vert}\tens b\t,\quad
\und\eps b=\eps
b,\quad \und Sb=g^{\vert b\vert}Sb}
\eqn{trans-any-univR}{\und\Delta^{\rm op} b=\sum b\t g^{-2\vert b\o\vert}\tens
g^{-\vert
b\t\vert}b\o,\quad \und\CR=\CR_{\Z_n'}^{-1}\sum\Ro g^{-\vert \Rt\vert}\tens
\Rt}
\end{corol}
\proof We apply the transmutation theorem, Theorem~5.1 and compute the form of
$B=B(\Z_n',H)$. Here $\Z_n'$ is the non-standard quantum group in (\ref{any-R})
with universal R-matrix $\CR_{\Z_n'}$. The action of $g$ on $H$ is in the
adjoint representation $g\la b=gbg^{-1}$ for $b\in \und H$ and
defines the degree of homogeneous elements by $g\la b=e^{2\pi\imath\vert
b\vert\over n}b$.
\endproof

\begin{corol} Let $H$ be a quantum group containing a group-like element $g$ of
order $2$ (so $g^2=1$ and $\Delta g=g\tens g$). Then $H$ has a corresponding
super-version $B$.
\end{corol}
\proof The formulae are as in (\ref{trans-any-hopf}) and
(\ref{trans-any-univR}) with $n=2$. \endproof

The first corollary was applied, for example to $H=u_q(g)$ at a root of unity
to simplify its structure. It leads to a new simpler form for its universal
R-matrix (by finding its anyonic universal R-matrix and working
back)\cite{Ma:any}. The second corollary was usefully applied in
\cite{MaPla:uni} to superise the non-standard quantum group associated to the
Alexander-Conway polynomial. In these examples, a sub-quantum group is used to
generate the braid statistics (braided category) in which the entire quantum
group is then viewed by transmutation. In the process its quasitriangular
structure or universal R-matrix becomes reduced because the part from the
subgroup is divided out (see Theorem~5.1). This means that the part
corresponding to the subgroup is made in some sense cocommutative.

\begin{corol}\cite{Ma:bra} Every strict quantum group (with universal R-matrix)
has a braided-group analogue $B(H,H)$ which is braided-cocommutative in the
sense that $\und\CR=1\tens 1$ and $\und\Deltaop=\und\Delta$. The latter is
\eqn{B(H)-cocom}{ \sum \Psi(b_{\und {(1)}}\tens Q\uo\la b_{\und
{(2)}})Q\ut=\sum b_{\und {(1)}}\tens b_{\und {(2)}}.}
We call $B(H,H)$ the {\em braided group of enveloping algebra type} associated
to $H$. It is also denoted by $\und H$\cite{Ma:bra}.
\end{corol}
\proof  Here we take the transmutation principle to its logical extreme and
view any quantum group $H$ in its own braided category $\Rep(H)$, by
$H\subseteq H$. This is a bit like using a metric to determine geodesic
co-ordinates. In that co-ordinate system the metric looks locally linear.
Likewise, in its own category (as a braided group) our original quantum group
looks braided-cocommutative. \endproof

This completely shifts then from one point of view (quantum=non-cocommutative
and bosonic object) to the other (classical=`cocommutative' but braided
object), and means that ordinary quantum group theory is contained in the
theory of braided-groups. On the other hand, not all braided-Hopf algebras are
obtained in this way. We shall encounter some that do not appear to come from
quantum groups in Chapter~7.

We have said in Theorem~5.1 that the resulting braided-Hopf algebra $B$ is
equivalent to the original one. The sense in which this is true is that spaces
and algebras etc on which $H$ act also become transmuted to corresponding ones
for $B$. Partly, this is obvious since $B=H$ as an algebra, so any
representation $V$ of $H$ is also a representation of $B$. The key point is
that $V$ is also acted upon by $H_1$ through the mapping $H_1\to H$. So the
action of $H$ is used in two ways, both to define the corresponding action of
$B$ and to define the `grading' of $V$ as an object in a braided category
$\Rep(H_1)$ (as explained at the end of Chapter~3). This extends the process of
transmutation to view a representation of $H$ also as a braided-representation
of $B$.

\begin{propos}\cite[Prop~3.2]{Ma:bos} If $C$ is an $H$-covariant algebra
($H$-module algebra) in the sense of (\ref{H-modalg}) then its transmutation is
a $B$-module algebra in the sense of (\ref{C-modalg}). Here the transmutation
does not change the action, but simply views it in the braided category.
\end{propos}

Thus covariant algebras for quantum groups as in (\ref{H-modalg}) become
braided-covariant algebras for their transmuted (super, anyonic or generally
braided) versions. For example, the adjoint action of $H$ on itself transmutes
to the braided-adjoint action of $B=B(H,H)$ on itself as studied in Chapter~4.
Moreover, it means that $B(H,H)$ is braided-cocommutative with respect to ${\rm
Ad}$ as promised there. Indeed

\begin{propos}\cite{Ma:bra}\cite{Ma:tra} For $B(H_1,H)$ the $\Deltaop$ behaves
like an opposite coproduct on all braided-representations that arise from
transmutation. In particular, $B(H,H)$ is cocommutative in the sense of
(\ref{V-cocom}) with respect to all such braided-representations that arise
from transmutation.
\end{propos}
\proof Writing the braids in (\ref{V-cocom}) in terms of the universal R-matrix
$\CR$ as explained in (\ref{Psi-R}) we see that the condition for all $V$ is
implied by (and essentially equivalent to) the intrinsic
braided-cocommutativity formula (\ref{B(H)-cocom}) in Corollary~5.4. \endproof

To conclude this topic, we make some general remarks about how  the formulae in
Theorem~5.1 were obtained. The idea is that not only do quantum groups generate
braided categories as in Chapter~3, but conversely there are classical theorems
that any braided category for which the objects can be strongly identified with
vector spaces, is essentially of the form $\Rep(H)$ for some quantum group $H$.
Strongly identified means mathematically a monoidal functor $\CC\to {\rm
VectorSpaces}$. (If we are given only a weaker identification, we may obtain
only some kind of weaker quantum group as actually happens in realistic
topological quantum field theories). The idea behind transmutation is that the
category $\CC$ could be targeted wherever we like, for example to super-vector
spaces or to another braided category $\CV$. Then the generalised
reconstruction theorem\cite{Ma:rec} would give a quantum group $\Aut(\CC,\CV)$
living not in the usual category of VectorSpaces, but in our chosen $\CV$. Thus
is how we can shift or transmute the category in which an algebraic structure
lives, by targeting its category of representations elsewhere. Some of the
philosophy behind this is developed in \cite{Ma:pri}. Moreover, the
braided-quasitriangular structure is given by the ratio of the braidings in
$\CC$ and $\CV$ so that the case $\aut(\CC)=\aut(\CC,\CC)$ is
braided-cocommutative. For example, given $H_1\to H$, we have such a functor
$\Rep(H)\to \Rep(H_1)$ and can reconstruct a braided-quantum group living in
$\Rep(H_1)$ and in the case $H_1=H$ it is braided-cocommutative. The detailed
structure is computed in \cite{Ma:bra}\cite{Ma:tra} with results as above.

\bigskip

What about going the other way, from braided groups to quantum groups? Not all
braided groups are of the type coming from transmutation, so we cannot simply
apply the above formulae in reverse. Here the unification of statistics and
covariance in Lemma~3.2 comes to our aid. For if $B$ is a braided-Hopf algebra
in a category of the form $\Rep(H)$, then to say that $H$ lives in the category
is to say equivalently that $B$ is fully $H$-covariant (its product and
coproduct, antipode etc all commute with the action of $H$). Now, when a
structure such as a quantum group acts covariantly on some other structure
(algebra, coalgebra etc), we can make a semidirect product.

\begin{theorem}\cite[Thm~4.1]{Ma:bos} Suppose that $B$ is a braided-Hopf
algebra living in a braided category of the form $\Rep(H)$ with action $\la$ of
$H$. Then there is also an induced coaction $\beta:B\to H\tens B$ of $H$ on the
coalgebra of $B$ given by  $\beta(b)=\CR_{21}\la b$. Here $\CR$ is the
universal R-matrix of $H$ and acts here with its first factor on $b$.
The cross product algebra by $\la$ and cross coproduct coalgebra by $\beta$ fit
together to form an ordinary Hopf algebra ${\rm bos}(B)=B\cocross H$.
\end{theorem}
\proof Perhaps the simplest proof is by direct computation using elementary
Hopf algebra techniques and the axioms in (\ref{univR}). For example, the
conversion of modules to comodules by $\CR$ is in \cite{Ma:dou}. The original
conceptual proof in \cite{Ma:bos} can be sketched as follows. Because $B$ is
acted upon by $H$, it becomes automatically (by transmutation) a
$B(H,H)$-module algebra in the braided category. Hence we can make a braided
cross-product using the formulae as shown on the top left in (\ref{C-cross})
with $B$ in the role of $C$ and $B(H,H)$ in the role of $B$ there. Because
$B(H,H)$ is cocommutative with respect to $B$, the result is a braided-Hopf
algebra $B\cocross B(H,H)$ with the braided tensor product coalgebra structure
(in analogy with the standard situation for usual group or Lie algebra cross
products). We can then recognise the result as that obtained by transmutation
from some ordinary quantum group ${\rm bos}(B)$ characterised by $B(H,{\rm
bos}(B))=B\cocross B(H,H)$. \endproof

The resulting bosonised Hopf algebra can be built on $B\tens H$ equipped with a
cross product algebra and coalgebra. The algebra is generated by $B,H$ with
cross relations  and crossed coproduct
\eqn{bos-structure}{\sum h\o b Sh\t=h\la b,\quad \Delta b=\sum b\Bo\CR\ut\tens
\CR\uo\la b\Bt}
where $b\equiv (b\tens 1$ and $h\equiv (1\tens h)$ and $H$ is a sub-Hopf
algebra. This ${\rm bos}(B)$ is equivalent to the original $B$ in the sense
that its ordinary representations correspond to the braided-representations of
$B$\cite{Ma:bos}.

\begin{corol} Every super-group (or super-quantum group) can be bosonised to an
equivalent ordinary quantum group. It consists of adjoining an element $g$ with
relations $g^2=1$, $gb=(-1)^{|b|}bg$ and
\eqn{bos-sup}{\Delta g=g\tens g,\ \Delta b=\sum b\Bo g^{|b\Bt|}\tens b\Bt,\quad
S b=g^{-|b|}\und S b,\quad \CR=\CR_{\Z_2'}\sum \und\CR\uo g^{|\und\CR\ut|}\tens
\und\CR\ut.}
\end{corol}
\proof We have seen in Proposition~3.1 that the category of super-vector spaces
is of the required form, with $H=\Z_2'$.
Here the super-representations of the original super-(quantum)-group are in
one-to-one correspondence with the usual representations of the bosonised
algebra. We have written the formulae in a way that works also in the anyonic
case with 2 replaced by $n$ and $(-1)$ by $e^{2\pi\imath\over n}$ for the Hopf
algebra structure.
\endproof

This means that the theory of super-Lie algebras (and likewise for colour-Lie
algebras\cite{Sch:gen}, anyonic quantum groups etc) is in a certain sense
redundant -- we could have worked with their bosonized ordinary quantum groups.
This is especially true in the super or colour case where there is no braiding
to complicate the picture. An application to physics is in
\cite{MacMa:spe}\cite{MacMa:str} where we observe that the spectrum generating
algebra of the harmonic oscillator is more naturally a quantum group than the
usual $osp(1|2)$.

\bigskip

This completes our lightning survey of transmutation and bosonization. As we
noted at the end of Chapter~4, all our constructions have left-right reflected
versions (like the above but with right-modules rather than left-modules) and
dual versions (with left-comodules or after reflection, right comodules). The
right-comodule theory is obtained by obtaining all diagram proofs above (and
the diagram-proofs in a diagrammatic form of the transmutation theorem)
up-side-down. We conclude the chapter by summarising the formulae above in this
dual form.

Firstly, instead of working with quasitriangular Hopf algebras $H$ to generate
the braid-statistics as in Chapter~3, we must work with dual-quasitriangular
Hopf algebras $(A,\CR)$ where $\CR:A\tens A\to \C$ is the dual-quasitriangular
structure and obeys
\eqn{dual-univR}{\CR(ab\tens c)=\sum
\CR(a\tens c\o)\CR(b\tens c\t),\ \CR(a\tens bc)=\sum
\CR(a\o\tens c)\CR(a\t\tens b)}
\eqn{dual-univR2}{ \sum b\o a\o \CR(a\t\tens b\t)=\sum \CR(a\o\tens b\o) a\t
b\t.}
and is invertible in a certain convolution algebra. This is just the dual of
(\ref{univR}). Now it is the category of $A$-comodules rather than $H$-modules
that is braided. So we will work in the braided category $\CC={\rm CoRep}(A)$.
The braiding $\Psi_{V,W}$ is given by applying the coactions to $V,W$ and
evaluating with $\CR$ on the $A\tens A$ that results. As the same time one
makes the usual permutation $P$ on the remaining $V\tens W$. This is the
analogue of (\ref{Psi-R}). The unification of statistics and covariance in
Lemma~3.2 now reads that the notions of and algebra in ${\rm CoRep}(A)$, and of
an $A$-comodule algebra, coincide.

The transmutation theory also has an analogue for $A\to A_1$ a Hopf algebra map
where $A_1$ is dual quasitriangular. Then $A$ has the additional structure of a
braided-group $B(A,A_1)$ with
\eqn{B(A)}{ B(A,A_1)=\cases{A&{\rm as\ a\ coalgebra}\cr  \und\cdot,\und S&{\rm
modified\ product\ and\ antipode}}}
For brevity we focus on the identity map (the general case is strictly
analogous), then the modified structures take the explicit
form\cite{Ma:bg}\cite{Ma:eul}
\eqn{B(A)-hopf}{a\und\cdot b=\sum a\t b\t \CR((S  a\o)a\th\tens S  b\o),\quad
\und{S }a=\sum S  a\t\CR((S ^2a\th)S  a\o\tens a_{(4)})}
and live in the braided category ${\rm CoRep}(A)$ by the right quantum-adjoint
coaction $\beta(a)={\rm Ad}_R(a)=\sum a\t\tens (Sa\o)a\th$. This time $B(A,A)$
is braided-commutative in the sense of (\ref{V-cocom}) turned up-side-down, for
all comodules $V$ that come from transmutation of comodules of $A$. This
reduces to an intrinsic form of commutativity dual to (\ref{B(H)-cocom}), and
from what we have said so far, it can be computed explicitly as
\eqn{B(A)-com}{b\und \cdot a=\sum a\th\und\cdot b\th \CR(Sa\t\tens
b\o)\CR(a_{(4)}\tens b\t)\CR(b_{(5)}\tens Sa\o)\CR(b_{(4)}\tens a_{(5)}).}
We call $\und A=B(A,A)$ the {\em braided group of function algebra type
associated to $A$}. A direct proof that these formulae define a braided-Hopf
algebra as in (\ref{hopf-ax}) appears in \cite[Appendix]{Ma:bg}.

\begin{propos} If $A$ is dual to $H$ then the corresponding braided groups
$B(A,A)$ and $B(H,H)$ are dual in the sense $B(A,A)^*=B(H,H)$ in
(\ref{B-dual}).
\end{propos}
\proof Here we view $B(A,A)$ as living in $\Rep(H)$ by the quantum-coadjoint
action  defined by $<h\la a,g>=<a,\sum (S  h\o)gh\t>$ on test-elements $g\in
H$. Roughly speaking, a right $A$-comodule is the same thing (at least in the
finite-dimensional case) as a left $H$-module by evaluation. Thus we have two
braided-Hopf algebras $B(H,H)$ and $B(A,A)$ in $\Rep(H)$. From the duality
principle above they must be dual in the category. Explicitly, the duality is
given by $b\in B(H,H)$ mapping to a linear functional $<Sb,(\ )>$ on $B(A,A)$,
where $S$ is the usual antipode of $H$. See \cite{Ma:mec} for full details. The
reader should be warned that the categorical dual as in (\ref{B-dual}) reduces
in the bosonic case not to the usual dual but to the usual dual with opposite
product and coproduct (which is isomorphic to the usual dual via the antipode).
\endproof

It is these dual formulae for the braided-commutative braided-Hopf algebra
$B(A,A)$ that leads to the braided matrix examples that we began with in
Chapter~2. Thus if $A$ is a matrix quantum group obtained as quotient of the
FRT bialgebra $A(R)$\cite{FRT:lie} we know that $A(R)$ and (we suppose) $A$ is
dual-quasitriangular. This is true of course for the standard $R$-matrices
(since $U_q(g)$ has a universal R-matrix\cite{Dri}) but is also true for
general non-standard $R$-matrices as proven in \cite[Sec. 3]{Ma:qua}. The
dual-quasitriangular structure consists of
\eqn{A(R)-univR}{\CR(\vect_1\tens\vect_2)=R_{12}}
extended according to (\ref{dual-univR}). Also, in Theorem~5.1 and its dual
version one can transmute not only Hopf algebras but bialgebras. In particular,
we can have $B(R)=B(A(R),A)$ as a braided-bialgebra in the category of
$A$-comodules. The general formulae above become in this case that $B(R)$
coincides with $A(R)$ as a coalgebra (so has generators $\vecu$ say with the
matrix braided-coproduct (\ref{B(R)-coprod}) and is covariant under the right
quantum-adjoint coaction
\eqn{Ad-coact-mat}{\beta(u^i{}_j)=u^m{}_n \tens (St^i{}_m)t^n{}_j,\qquad{\rm
i.e.,}\quad \vecu\to\vect^{-1}\vecu\vect.}
Meanwhile, the braided-commutativity relations (\ref{B(A)-com}) reduce using
(\ref{A(R)-univR}) to a matrix equation of the form $uu=uu R^{-1}R\widetilde R
R$ with appropriate indices, see \cite{Ma:eul}\cite{Ma:exa}. Putting two of the
$R$'s to the left (or rearranging (\ref{B(A)-com})) gives the
braided-commutativity relations (\ref{bra-com}) in the compact form used in
Chapter~2\cite{Ma:lin}. Finally, $\Psi$ computed from (\ref{Ad-coact-mat}) and
(\ref{A(R)-univR}) gives the braid-statistics (\ref{bra-mat}) used there.

This is how the braided-matrices in Chapter~2 were first introduced. We obtain
not only the results in Chapter~2 but also that they are related to usual
quantum matrices by transmutation. The formula for the modified product comes
out from (\ref{B(A)-hopf}) as \cite{Ma:lin}
\eqn{u-t-trans}{\begin{array}{rcl}u^i{}_j&=&t^i{}_j\\  u^i{}_j
u^k{}_l&=&t^a{}_bt^d{}_lR^i{}_a{}^c{}_d \widetilde R^b{}_j{}^k{}_c \\ u^i{}_j
u^k{}_l u^m{}_n&=&  t^d{}_b t^s{}_u t^z{}_n R^i{}_a{}^p{}_q R^a{}_d{}^w{}_y
{\widetilde
R}^b{}_c{}^v{}_w {\widetilde R}^c{}_j{}^k{}_p R^q{}_s{}^y{}_z
{\widetilde R}^u{}_l{}^m{}_v \end{array}}
etc. Here the products on the left are in $B(R)$ and are related by
transmutation to the products on the right which are in $A(R)$. If we write
some or all of the $R$-matrices over to the left hand side we have equally well
the compact matrix form \cite{Ma:lin},
\eqn{t-u-trans}{\begin{array}{rcl}\vecu &=&\vect\\ R^{-1}_{12}\vecu_1
R_{12}\vecu_2& = &\vect_1\vect_2 \\
R_{23}^{-1} R_{13}^{-1}R_{12}^{-1}\vecu_1 R_{12}\vecu_2  R_{13}R_{23}\vecu_3&=&
\vect_1\vect_2\vect_3 \end{array}}
etc. This is just a rearrangement of (\ref{u-t-trans}) or our universal formula
(\ref{B(A)-hopf}). For the transmuted product of multiple strings the universal
formula from (\ref{B(A)-hopf}) involves a kind of partition function made from
products of $R$ to transmute the bosonic $A(R)$ to the braided $B(R)$
\cite{Ma:lin}.

Because of these transmutation formulae, braided groups are a useful tool (a
kind of conceptual `co-ordinates') for doing ordinary quantum group
computations. On the braided side they can be much simpler or more like the
group case because of the braided-commutativity of $B(R)$ (which make it behave
more like a classical group as we have seen in Chapter~4), after which formulae
can be converted back by applying the transmutation formulae
(\ref{u-t-trans})-(\ref{t-u-trans}) in the reverse point of view. This was the
reason given for introducing the transmutation process in
\cite{Ma:eul}\cite{Ma:bg}.

\section{Applications to Quantum Groups}

In this chapter we mention a few selected applications of the above theory to
quantum groups. The first makes use of the unification of statistics and
covariance in Chapter~3 to study quantum-group covariant systems. The other two
make use of the transmutation theory in the last chapter. There are many more
applications of braided groups as mentioned in the introduction.

\subsection{Quantum-Covariant Spin Chains and Exchange Algebras}

 The braided-tensor product construction in Chapter~4 in the form of
Corollary~4.2 has many applications in physics and $q$-deformed physics, for it
tells us precisely how to combine two quantum group covariant systems in a
quantum-group covariant way. One has to use the braiding $\Psi$ given by the
action of the universal R-matrix of the quantum group.

In particular, given one quantum-group covariant algebra $C$ we can repeatedly
take the braided-tensor product $C^{\und\tens^N}$ for the corresponding N-body
system. Here we regard each $C$ as the quantum algebra of observables of a
single particle or site and $C^{\und\tens^N}$ as the system for a chain of $N$
such particles with a natural interaction or statistics designed to maintain
quantum-group covariance. Building up quantum-spin chains in this way ensures
that they are manifestly quantum group covariant. The importance of quantum
covariance for quantum spin chains is well-known in physics, see for example
\cite{Kul:alg}.

As an example, one can take for $C$ the braided-Heisenberg group corresponding
to the quantum Heisenberg group $U_q(h)$ of \cite{CGST:hei}. As an algebra it
is basically the usual harmonic oscillator with $\hbar$ viewed as a central
operator rather than a constant, and has braid statistics $\Psi$ which tend to
trivial bosonic statistics as $q\to 1$. This model has been worked out in
complete detail by W.K. Baskerville in \cite{BasMa:bra}. At first sight it
might seem that the system of $N$ braided-harmonic oscillators with
braid-statistics between them, would become horribly tangled. However, it is
found instead that the system is in fact non-trivially isomorphic as an algebra
to the unbraided system of $N$ commuting copies of $U_q(h)$. This allows
constructions on the braided-side, which are manifestly covariant, to be mapped
over to the unbraided side. In particular, $U_q(h)$ acts covariantly on the
system and the action of one of its generators (the number operator) recovers
the usual free particle evolution. One can also add covariant interaction terms
to the Hamiltonian using these methods. We refer to \cite{BasMa:bra} for all
these results.

The general principle here is that in general $B(H,H)^{\und\tens^n}\isom
H^{\tens^n}$ as an algebra, using the isomorphism that connects $\Delta$ to
$\und\Delta$ or $\rho$ to $\und\CR$ in the transmutation formulae
(\ref{B(H)-hopf})-(\ref{B(H)-univR}). Recall that the universal R-matrix is
needed here in a non-trivial way. One can apply it also, for example to
$BM_q(1,1)^{\und\tens^N}$ in (\ref{bm1|1a})-(\ref{bm1|1stat}). This is a system
with bose-fermi like statistics (and in the limit $q\to 1$ becomes an N-fold
product of super-matrices). The right hand side on the other hand is the usual
tensor product of a bosonic quantum group associated to the relevant
$R$-matrix.

We can also take N-fold or infinite braided-tensor products such as
$V^*(R')^{\und\tens ^\infty}$\cite{Ma:inf} of the braided-(co)vectors $V^*(R')$
and $V(R')$ (which do not come from transmutation). As well as being thought of
a quantum-spin chains, we can also think of these in terms of braided-geometry.
There is a braided vector space at each site, and the independent sites have
braid-statistics. Thus they should be thought of precisely as
braided-vector-fields (or sections of a braided-vector bundle) in one space
dimension (in a lattice approximation).
For example, the structure of $V^*(R')^{\und\tens ^\infty}$ comes out as
follows. Let $x_i(m)$ where $m\in \N$ denote the generators of the copy of
$V^*(R')$ in the $m$'th position in the infinite (or finite) braided tensor
product and 1 elsewhere. If $m<n$ then $x_i(m)x_j(n)=\cdots\tens
x_i\tens\cdots\tens x_j\cdots$ while if $m>n$ we must use $\Psi$ to take
$\cdots \tens x_i\tens \cdots$ past $\cdots\tens x_j\tens\cdots $ when we
multiply as in (\ref{tensprod}). Here the braiding is $\Psi(x_i\tens
x_j)=x_l\tens x_k R^k{}_i{}^l{}_j$ as in (\ref{bra-vec}) so we obtain
$\cdots\tens x_l\tens\cdots\tens x_k\tens\cdots
R^k{}_i{}^l{}_j=x_l(n)x_k(m)R^k{}_i{}^l{}_j$. Thus the braided tensor product
is a kind of `ordered
product' of the $x_i$ \cite{Ma:inf}. From this we can write
$V^*(R')^{\und\tens ^\infty}$ as generated by $\{x_i(m)\}$ with relations
\eqn{inftens-rel}{\vecx_1(m)\vecx_2(n)=\vecx_2(n)\vecx_1(m)R(n-m)}
where
\eqn{R(n-m)}{ R(n-m)=\cases{R& if\ $m>n$\cr R'& if\ $m=n$\cr
R^{-1}_{21}& if\ $m<n$}}
obeys the parametrized Yang-Baxter equations
\eqn{param-YBE}{
R_{12}(m-n)R_{13}(m)R_{23}(n)=R_{23}(n)R_{13}(m)R_{12}(m-n)}
with discrete spectral parameter in the sense $m,n\in \N$, cf\cite{Ma:seq}.

These are relations of precisely the form that are noted for the exchange
algebra in conformal field theory and in 2D quantum gravity. See for
example\cite{Ger:str} and later works. In this context there are fields
$\xi(\sigma)$ obeying
$\xi_1(\sigma)\xi_2(\sigma')=\xi_2(\sigma')\xi(\sigma) R(\sigma'-\sigma)$ where
$R(\sigma'-\sigma)=\cases{R& if\ $\sigma>\sigma'$\cr R^{-1}_{12}& if\
$\sigma<\sigma'$}$. Here $\sigma,\sigma'\in [0,1]$ are continuous and one does
not worry about the case of equality. We see then that this exchange algebra
has the mathematical structure of a continuous braided-tensor product of
braided-covectors or braided-vectors. This is striking for several reasons.
Firstly, we see at once, by construction, that the fields $\xi_i$ generate a
quantum-group-covariant algebra (because they are built up using the braided
tensor product). The action or coaction is pointwise. Secondly, because
$V^*(R')$ has a braided Hopf algebra structure corresponding to
braided-covector addition as explained in Chapter~2, we have in the braided
tensor product or (by iterating) in any braided tensor power,  a realization
$\und\Delta^N:V^*(R')\to V^*(R')^{\und\tens N}$. In the present case this comes
out from (\ref{V^*(R)-coprod}) as the realization
\eqn{int-exch}{x_i=\sum_m x_i(m),\qquad {\rm i.e.,} \quad \xi_i=\int d\sigma\,
\xi_i(\sigma).}
Moreover, there are all sorts of other braided-linear algebra constructions
that one can make. For example, the pointwise-version of (\ref{proj-mat}) means
that using $\xi$ and its conjugate fields one should be able to realise
rank-one braided matrices. As we shall see in Chapter~7.1 this is not a quantum
group but, in the $SL_q(2)$ case, the degenerate Sklyanin algebra.  Indeed,
because the relations of $V^*(R')$ are some kind of (braided)-commutativity
relation, what we gain by our approach is a picture of this version of the
exchange algebra as describing a classical but braid-statistical (like
fermionic) vector-field, and yet equivalent to its usual picture as describing
a quantum field. This is like the transmutation principle above and is a step
to a formulation of the geometrical structure of such quantum field theories.

Note that one can do the same thing with $V^*(\lambda R)$ where $\lambda$ is a
suitable constant and this is the usual Zamolodchikov algebra. This is also a
covariant algebra which can be tensor-producted (this was the actual point of
view in \cite{Ma:inf}) but because $V^*(\lambda R)$ is not in general a
braided-Hopf algebra, we do not have the full picture as above. It appears to
coincide only in the Hecke case where $R'=\lambda R$. Thus it appears that the
right generalization of the Zamolodchikov and exchange algebras to general
non-Hecke R-matrices (such as the $SO_q(1,3)$ $R$-matrix) is with $R'$ and not
$\lambda R$ if we wish to have a full picture.

\subsection{Self-Duality and Factorizable Quantum Groups}

In Chapter~5 we computed the braided groups and braided-matrices arising from
transmutation of quantum function algebras such as $A(R)$. We can also compute
the braided groups arising from the standard quantum groups $U_q(g)$ in
Theorem~5.1. This was done in \cite{Ma:skl} and the result is as follows. For
$U_q(g)$ we take the generators in `FRT' form\cite{FRT:lie} as
$\vecl=\{l^\pm{}^i{}_j\}$, and write $L=\vecl^+ S\vecl^-$. Then the
braided-coproduct for $B(U_q(g),U_q(g))$ in Theorem~5.1 comes out
as\cite{Ma:skl}
\eqn{L-coprod}{\und\Delta L=L\tens L.}
Although the generators $L$ for $U_q(g)$ have been useful in the past, this
(\ref{L-coprod}) was not discussed or used  because it does not form an
ordinary Hopf algebra (the ordinary coproduct is different). One needs the
theory of braided-Hopf algebras to appreciate it.

Next, a quantum group $H$ is called {\em factorizable}\cite{ResSem:mat} if
$Q=\CR_{21}\CR_{12}$ is non-degenerate in a certain sense. We call $Q$ the {\em
quantum Killing form}. The condition is that the map $Q:A\to H$ given by $a\to
(a\tens\id)(Q)$ is a linear isomorphism, where $A$ is the dual of $H$. The
usual quantum groups $U_q(g)$ with their suitable quantum function algebras
$G_q$ as dual are factorizable  (at least up to suitable formulation of
generators). So we have two quantum groups related by $Q$ as linear spaces.
Each has a braided version as in Proposition~5.9. Remarkably, one can show by
explicit computation that\cite{Ma:skl}
\eqn{Q-isom}{Q:B(A,A)\to B(H,H)}
as a homomorphism of braided-Hopf algebras, and an isomorphism in the
factorizable case. See also \cite{LyuMa:bra}. That $Q$ is a morphism says that
it is quantum-$\Ad$ invariant, which is indeed a property of the quantum
Killing form, in analogy with the usual (inverse) Killing form in $g\tens g$
\eqn{K-killing}{ K^{-1}:g^*\to g.}
What the braided theory tells us is much more than this semiclassical limit: it
tells us that the homomorphism or isomorphism is not only as linear spaces (as
in the semiclassical case) but as entire braided-groups, mapping their algebras
and braided-coproducts etc on onto the other. This, the semiclassical notion of
semisimplicity of $g$ is a remnant of something deeper: the {\em self-duality
of the braided-groups}  $B(A,A)\isom B(H,H)$. Although conceptually very
different (one is of function algebra type, and the other of enveloping algebra
type) they become the same in the factorizable case.
For $U_q(g)$ one has $Q=1+2\hbar K^{-1}+O(\hbar^2)$. Here $\hbar$ is a
deformation parameter rather than physical Planck's constant.

In the case of $U_q(g)$ we have seen that $B(A,A)$ is a quotient of $B(R)$ (by
braided determinants etc) and has generators $\vecu$ as in Chapter~2. The map
$Q$ then comes out as\cite[Sec. 2]{Ma:skl}
\eqn{Q-B(R)}{Q(\vecu)=<\CR_{21}\CR_{12},\vect\tens\id>=\vecl^+ S\vecl^-=L.}
This then explains facts about $U_q(g)$ that have been found in other ways.
Firstly, it is well-known that {\em some} of the relations of $U_q(g)$ take the
form $R_{21}L_1 R_{12} L_2=L_2 R_{21} L_1 R_{12}$ and we see from our point of
view that this is a consequence of the self-duality expressed in
(\ref{Q-B(R)}). Also, the bosonic elements $c_k$ from (\ref{bos})  recover the
Casimirs of $U_q(g)$ first found in \cite{FRT:lie} as $Q(c_k)$. Note that
(\ref{Q-isom}) and (\ref{Q-B(R)}) are a purely quantum phenomenon (with the
above semiclassical remnant) in the sense that the map $Q$ is trivial (or
$Q^{-1}$ is rather singular) at $q=1$. Thus we are recovering results about
$U_q(g)$ by working strictly at generic $q\ne 1$.

All this has many concrete applications. We mention here only one, where we
showed that Drinfeld's quantum double $D(U_q(g))$ (of some interest in physics
for various reasons) is in fact isomorphic to a semidirect product
$U_q(g)\cocross U_q(g)$ by the quantum adjoint action\cite{Ma:skl}. In this new
form of the quantum double, the generators are $\vecl^\pm$ and ${\bf m}^\pm$
say (two copies of $U_q(g)$) with cross relations\cite[Corol~4.3]{Ma:skl}
\eqn{double}{ R_{12}l_2^+M_1=M_1 R_{12}l_2^+,\quad R_{21}^{-1}l_2^-
M_1=M_1R_{21}^{-1}l_2^-}
where $M={\bf m}^+S{\bf m}^-$. The coalgebra is also a semidirect coproduct.
Mathematically, this form of the quantum double is obtained as nothing other
than an example of the bosonization Theorem~5.7 applied to $B(U_q(g),U_q(g))$.
The quantum adjoint action here can also be computed as\cite[Prop~2.4]{Ma:skl}
\eqn{q-ad-act-mat}{L^i{}_j\la L^k{}_l=\widetilde{R}{}^a{}_{j}{}^{k}{}_b
R^{-1}{}^b{}_{d}{}^{i}{}_c Q^c{}_{a}{}^{e}{}_l L^d{}_e}
where $Q=R_{21}R_{12}$. This can be used to write the action of $L$ on $M$ for
the cross relations in $U_q(g)\cocross U_q(g)$ (in place of (\ref{double})), as
well as to define some kind of quantum or braided Lie bracket\cite{Ma:skl}.
We refer to \cite{Ma:skl} for the details of these results.

\subsection{Action-Angle Variables for Quantum Groups}

If we regard $A(R)$ and other quantum function algebras as analogous to some
kind of quantum algebra of observables (this is {\em not} their actual role in
physics in QISM), then it is natural to look for a complete commuting set of
observables. The braided theory above does precisely this. Let $R$ be a matrix
obeying the QYBE. It need not be standard but should be generic enough that the
second inverse $\widetilde R$ exists. Then we find
\eqn{action-angle}{ \{\alpha_k\ \in A(R)\},\qquad [\alpha_k,\alpha_l]=0,\qquad
\Ad_R(\alpha_k)=\alpha_k\tens 1.}
In the standard case there are as many algebraically independent $\alpha_k$
(roughly speaking) as the rank of the Lie algebra, so in some sense these
$\alpha_k$ are a `complete' set.

They can perhaps be called `quantum angle variables' in honour of the role of
action-angle variables in the theory of classical inverse scattering; one may
hope that they could prove correspondingly useful in quantum inverse
scattering.
So far, they have been used in \cite{BrzMa:bic} to define a `ring' of
bicovariant differential calculi for any matrix quantum group. Here any
polynomial function in the $\alpha_k$ defines some $\Ad$-invariant element, and
every $\Ad$-invariant element defines a differential calculus\cite{BrzMa:bic}.
This function is a new kind of `field' in physics, one that governs the choice
of differential structure. It is interesting to ask if there is some kind of
action-principle for such fields leading to the usual differential structures
assumed in physics as extrema.

This result follows very simply from the transmutation theory in Chapter~5.
Recall that this is some kind of non-linear transformation that maps
\eqn{trans-map}{{\rm qu.\ non-commutativity\ of\ }A(R)\ \to\ \cases{{\rm
braided-commutativity\ of\ }B(R)&\cr{\rm \&\ statistical\ non-commutativity}&}}
Under this mapping, the bosonic and central elements
$c_k=\trace\vartheta\vecu^k$ in (\ref{bos}) are the image of some elements
$\alpha_k$ in $A(R)$. From (\ref{u-t-trans}) one finds at once\cite{BrzMa:bic}
\eqn{alpha_k}{\begin{array}{rcl}\alpha_1 &=&\vartheta^i{}_jt^j{}_i\\
\alpha_2 &=&\vartheta^i{}_j R^j{}_k{}^m{}_n
t^k{}_l\vartheta^l{}_m t^n{}_i\\
\alpha_3 &=&\vartheta^i{}_j R^j{}_k{}^p{}_q R^k{}_l{}^w{}_y t^l{}_m {\widetilde
R}^m{}_n{}^v{}_w \vartheta^n{}_p R^q{}_s{}^y{}_z t^s{}_u
\vartheta^u{}_v t^z{}_i\end{array}}
etc. One can compute all the $\alpha_k$ in a similar way from (\ref{u-t-trans})
or (\ref{B(A)-hopf}). Here $\alpha_1$ is the quantum trace (which is well-known
to be $\Ad$-invariant) but the higher $\alpha_k$ are also useful and can be
used, for example to obtain quantum determinants and other expressions.

The proof that these $\alpha_k$ obey (\ref{action-angle}) is then easy. That
the $c_k$ are bosonic comes from the fact that they are $\Ad$-invariant under
the quantum-adjoint coaction. Here we identify the original dual quantum group
$A(R)$ with the braided group $B(R)$ as linear spaces as in (\ref{B(A)}). The
product is modified, but from (\ref{B(A)-hopf}) we see that if $\Ad_R(a)=a\tens
1$ (quantum-$\Ad$ invariant) then $a\und\cdot b= ab$. Hence $\alpha_k=c_k$ as
elements of a linear space and $\alpha_k \alpha_l=c_k\und\cdot c_l=c_l\und\cdot
c_k=\alpha_l\alpha_k$.
Here we used that the $c_k$ are central in $B(R)$. The $\alpha_k$ themselves
are not central but we see that they mutually commute as a remnant under the
inverse of (\ref{trans-map}) of the fact that $c_k$ are both bosonic and
central.

\section{Applications Beyond Quantum Groups}

In this final chapter we announce from the preprints \cite{Ma:skl}\cite{Ma:poi}
a few selected applications of the above theory to problems where
quantum-groups have been found to fail. When one begins to systematically
$q$-deform everything in physics one soon finds that quantum groups are not
enough, and that more naturally, everything acquires braid statistics. If the
$q$ is physical then these braid statistics are physical. If the $q$ is only a
regularization parameter before renormalization\cite{Ma:reg} then the
braid-statistics are an artifact of the regularization procedure. However, they
can still have physical consequences after renormalisation (which can now be
done elegantly using braided or $q$-analysis) and setting $q\to 1$. Roughly
speaking the role of $q$ here is like a systematic variant of `point splitting'
(see (\ref{tensprod}) and (\ref{tens-act})) with a choice of braid crossing or
inverse-braid crossing reflecting topologically distinct ways to set $q\to 1$.
This approach to regularization could be especially relevant to quantum gravity
where some formulations already make use of knots and loops.

\subsection{The Degenerate Sklyanin Algebra}

An important algebra in quantum inverse scattering is the Sklyanin
algebra\cite{Skl:alg}. It is the homomorphic image of
a bialgebra associated to the 8-vertex solution of the parametrized Yang-Baxter
equations (and hence its representations lead to ones of this bialgebra). It
has 3-parameters $J_{12}, J_{23}, J_{31}$ obeying one equation
$J_{12}+J_{23}+J_{31}+J_{12}J_{23}J_{31}=0$, and an elegant structure in terms
of four generators $S_0,S_1,S_2,S_3$. On the other hand, in spite of many
efforts, there has not been found any bialgebra
structure on the Sklyanin algebra itself. Even in the degenerate case where one
of the parameters vanishes, which is associated with the 6-vertex solution and
the quantum group $U_q(sl_2)$, the degenerate Sklyanin algebra itself does not
appear to be a usual bialgebra.

We have shown in \cite{Ma:skl} that the degenerate Sklyanin algebra, in a form
with suitable generators, is a braided-bialgebra. It turns out to be isomorphic
to the braided matrices $BM_q(2)$ in (\ref{bm2a})-(\ref{bm2stat}). This comes
out as the identification\cite{Ma:skl}
\eqn{skl-bm2}{ \pmatrix{a&b\cr c&d}=\pmatrix{K^2_+& q^{-\h}(q-q^{-1})K_+Y_-\cr
q^{-\h}(q-q^{-1})Y_+K_+& K_-^2+q^{-1}(q-q^{-1})^2Y_+Y_-}}
where the four generators of the degenerate Sklyanin algebra are
$Y_\pm=\h\sqrt{1-t^2}(S_1\pm\imath S_2)$ and $K_\pm=S_0\pm tS_3$. Here
$t=\sqrt{J_{23}}$ (a fixed square root) is the remaining free parameter in the
degenerate case and $q={1+t\over 1-t}$. In making this identification  we allow
$a, d-ca^{-1}b$ of $BM_q(2)$ to be invertible and have square roots.

What this means is that the degenerate Sklyanin algebra lives in some braided
category in the sense explained in Chapter~3. In our case the relevant category
is the braided category of $U_q(sl_2)$-modules. If we denote by
$X_\pm,q^{\pm{H\over 2}}$ the usual generators of $U_q(sl_2)$ in Jimbo's
conventions, then the action comes out explicitly as\cite{Ma:skl}.
\eqn{skl-act}{\begin{array}{rll} &q^{H\over 2}\la K_\pm=K_\pm &q^{H\over 2}\la
Y_\pm=q^{\pm 1}Y_\pm\\
&X_\pm\la K_+=(1-q^{\pm 1})Y_\pm &X_\pm\la K_-=(q^{\pm 1}-1)K_+^{-1}Y_\pm K_-\\
& X_+\la Y_+=(1-q^{-1})Y_+^2 K_+^{-1} &X_+\la Y_-=K_+^{-1}(Y_+Y_--qY_-Y_+)\\
&X_-\la Y_-=(1-q)Y_-^2 K_+^{-1} &X_-\la
Y_+=K_+^{-1}(Y_-Y_+-q^{-1}Y_+Y_-).\end{array}}

The braid statistics and braided-coproduct for the degenerate Sklanin algebra
can likewise be worked out explicitly from (\ref{bm2stat}) (or from
(\ref{Psi-R})) and (\ref{B(R)-coprod}) respectively. The results on the
individual generators themselves rather than products involve infinite
power-series (because the universal R-matrix for $U_q(sl_2)$ is an infinite
power-series). An alternative is to realise that, more precisely, it is the
subalgebra of the degenerate Sklyanin algebra generated by $a,b,c,d$ in
(\ref{skl-bm2}) that is $BM_q(2)$ and has finite braiding and coproduct.

Either way (working with formal power-series or with the subalgebra) we have a
braided-bialgebra. Hence, for example, we can make braided-tensor products of
braided-representations as in (\ref{tens-act}). Moreover, we have the
braided-cocommutativity properties as in (\ref{V-cocom}) (for suitable
representations) and (\ref{comtens}) for their tensor product. It might be
thought that this braided-bialgebra is not much different from $U_q(sl_2)$
(because it is known that adding one more relation to the degenerate Sklyanin
algebra gives the algebra of $U_q(sl_2)$). But this is not so. For example, in
our braided-matrix description this additional relation turns out to be
$BDET(\vecu)=1$ as in  (\ref{bm2b-det}) and gives $U_q(sl_2)$ via the
self-duality (\ref{Q-B(R)})\cite{Ma:skl}. By contrast, the realization
(\ref{proj-mat}) of $BM_q(2)$ in terms of braided-vectors and
braided-covectors, and hence the corresponding realization of the degenerate
Sklyanin algebra, has $BDET(\vecu)=0$. So, such realizations of the degenerate
Sklyanin algebra, and associated representations, are far removed from the
usual realizations and representations of $U_q(sl_2)$. We refer to
\cite{Ma:skl} for the details.

\subsection{q-Homogeneous Spaces and q-Poincar\'e Group}

Among the several general results about braided groups in \cite{Ma:skl} we note
here one that says that quantum-homogeneous spaces are, in general, braided
groups.

Recall that if we have Lie algebras $h\subset g$ in a strong sense then
$g=m\cocross h$ is a semidirect sum Lie algebra. This happens for example, for
the Poincar\'e group Lie algebra $p\supset so(1,3)$. Likewise, if we have an
inclusion of ordinary Hopf algebras $H\subset H_1$ which is a strong one in the
sense that there is a Hopf algebra map $H_1\to H$ covering the inclusion then a
theorem in \cite[Prop. A.2]{Ma:skl} asserts that $H_1=B\cocross H$ as a
semidirect product and coproduct, where $B$ is a braided-Hopf algebra. It lives
in the braided category of $D(H)$-modules where $D(H)$ is Drinfeld's quantum
double construction. If $H$ is a strict quantum group (with universal R-matrix)
then the braided-category of $H$-modules is contained in that of
$D(H)$-modules, and $B$ often lives in this smaller category.
Either way, we see that quantum groups are not enough -- even in nice cases the
homogeneous spaces associated to inclusions of quantum groups force us to
braided groups.

On these very general grounds, we know for example that if we can find some
$q$-Poincar\'e group $U_q(p)$ strongly containing the known q-Lorentz group
$U_q(so(1,3))$ dual to $SO_q(1,3)$ in \cite{CWSSW:lor}, then $U_q(p)\isom
B\cocross U_q(so(1,3))$ for some braided-group $B$ in the role of Minkowski
space.

We can also use the theorem mentioned above for quantum function algebras
rather than quantum enveloping algebras. Now the usual inclusion of the Lorentz
group in the Poincar\'e group becomes a projection $P_q\to SO_q(1,3)$ where
$P_q$ is our supposed quantum function algebra for our quantum Poincar\'e
group. If we suppose that this is strong it in the sense that it covers an
inclusion $SO_q(1,3)\subset P_q$ of Hopf algebras, then we again have on
general grounds that $P_q\isom B\cocross SO_q(1,3)$. Here $B$ is a braided
group of `functions' on an analogue of Minkowski space.

Armed with this general picture, we have been motivated in \cite{Ma:poi} to
construct $P_q$ by beginning with a suitable candidate for $B$, since we know
from the above that it should be a braided-group version of $q$-Minkowski
space. Thus, we take $B=V^*(R')$ where $R$ is the q-Lorentz group R-matrix
(such as the one for $SO_q(1,3)$ in \cite{CWSSW:lor}). We have seen in
Chapter~2 that this has a braided-covector addition law (\ref{bra-covec}), so
we can add  $q$-Minkowski position vectors provided we treat them with braid
statistics. We denote $q$-Minkowski space with this braided addition law by
$\R^{1,3}_q$.

The first step is to identify more precisely the quantum group under which
$V^*(R')$ (and $V(R')$) are covariant. The simplest possibility would be to
take for this the bialgebra $A(R)$, which becomes dual-quasitriangular as
explained in (\ref{A(R)-univR}). Its category of comodules is braided and it
indeed coacts from the right on $V^*(R')$ by $\vecx\to\vecx\vect$. The induced
braiding is
\eqn{A(R)-covec-psi}{\Psi(\vecx_1\tens\vecx_2)=\vecx_2\tens\vecx_1
\CR(\vect_1\tens\vect_2)=\vecx_2\tens \vecx_1 R_{12}}
as used in (\ref{bra-covec}) and Chapter~6.1.

While this works for $A(R)$, this $\CR$ defined by (\ref{A(R)-univR}) needs $R$
to be correctly normalised if it is to descend to the relevant quantum group
$A$ (such as $SO_q(1,3)$). On the other hand, the normalization of $R$ in
(\ref{bra-covec}) is fixed by (\ref{R'}) and is not usually this quantum group
normalization. This forces us to extend $A$ as follows\cite{Ma:poi}. We
normalise $R$ according to (\ref{R'}) and let $\lambda R$ be the quantum group
normalization. The extension $\widetilde A$ is then given by adjoining to $A$ a
single invertible and commuting generator $g$, with coproduct $\Delta g=g\tens
g$ and dual-quasitriangular structure $\CR(g\tens g)=\lambda^{-1}$.

This $\widetilde A$ also coacts from the right on $V^*(R')$ and $V(R')$ in
Chapter~2, by
\eqn{vec-covec-cov}{\vecx\to \vecx\vect g,\qquad \vecv\to
g^{-1}\vect^{-1}\vecv}
and gives the correct braiding $\Psi$. In this way, these braided-vectors and
covectors live in the braided category of $\widetilde A$-comodules. We are
using here (in a dual form) the unification of statistics and covariance in
Chapter~3. In the present setting it means that $V^*(R')\to V^*(R')\tens
\widetilde A$ etc are comodule algebras (they are algebra homomorphisms, so the
algebra $V^*(R')$ is realised in $V^*(R')\tens \widetilde A$ by
(\ref{vec-covec-cov}) and similarly for $V(R')$). This means that
\eqn{vec-covec-comodalg}{\vecx_1\vecx_2\cdots\vecx_n\to
\vecx_1\vecx_2\cdots\vecx_n \vect_1\cdots\vect_n g^n}
so that we see that $g$ measures the degree of a polynomial in the $x_i$. It
coacts according to the physical scaling dimension.

Not only is the algebra of $V^*(R')$ respected by the coaction of $\widetilde
A$, but so are all the other braided-group maps, such as the braided-coproduct
corresponding to covector addition. This means that we can form a right-handed
semidirect product and coproduct algebra $\widetilde A\cross V^*(R')$ by a dual
version of the bosonization Theorem~5.7. For example, with an $SO_q(1,3)$
R-matrix we have $P_q=\widetilde{SO_q(1,3)}\cross {\R^{1,3}_q}$ as an ordinary
Hopf algebra. Explicitly, it consists of the generators of the $q$-Minkowski
space ${\R^{1,3}_q}$, which we now denote by $\{p_i\}$ (since we think of them
as momentum) and the generators $g,\vect$ of the extended quantum rotation
group, and relations and coproduct\cite{Ma:poi}
\eqn{poincare}{\vecp g=\lambda^{-1}g\vecp,\
\vecp_1\vect_2=\lambda\vect_2\vecp_1R_{12},\quad \Delta \vecp=\vecp\tens \vect
g+1\tens\vecp,\quad \Delta \vect=\vect\tens\vect, \ \Delta g=g\tens g.}
Relations of this form were first proposed as a kind of $q$-Poincar\'e group in
\cite{SWW:inh}, but we see now their structure as a semidirect product of a
momentum braided-covector space by quantum rotations. Its dual is a quantum
enveloping algebra also of the semidirect product form $U_q(p)=B\cocross
\widetilde H$.

Finally, because all our constructions are covariant, it is not hard to see
that this $q$-Poincar\'e group (\ref{poincare}) coacts from the right on
another copy of ${\R^{1,3}_q}$, which we think of as $q$-spacetime,
by\cite{Ma:poi}
\eqn{poin-mink-act}{\vecx\to \vecx\vect g+ \vecp}
The $q$-Minkowski spacetime is fully covariant under this (the right hand side
obeys the same algebra etc).
We defer to \cite{Ma:poi} for the details.

\subsection{Braided-Differential Calculus}

As a final application of braided-groups, we mention some results about
braided-differential calculus. The idea is to develop these strictly in analogy
with super-differentials. Thus, they obey what we call a {\em braided Leibniz
rule} and generate braided-translations as in (\ref{bra-covec}).

We begin with the simplest case, the {\em braided line}\cite{Ma:csta}. This has
a single generator $x$ with no relations. So the algebra $B$ is just
polynomials in one variable. To this, we add braid statistics and
braided-coproduct
\eqn{bra-line}{\Psi(x\tens x)=q^2 x\tens x,\qquad \und\Delta x=x\tens 1+1\tens
x}
In the language of Chapter~2 this means that an independent copy has braid
statistics $x'x=q^2 xx'$ and in this braided tensor product we realise $x+x'$.

It is very natural to differentiate this braided-addition law to obtain the
braided-differential\cite{Ma:poi}
\eqn{del-line}{\del_q f(x)=(f(x+\eps')-f(x))\eps'^{-1}|_{\eps'=0}}
where we mean here the linear part in $\eps'$ of $f(x+\eps')-f(x)$ and where
$\eps'$ has the braid-statistics relative to $x$. Using the $q$-binomial
theorem one finds
\eqn{del-leib}{\del_q x^m=[m]_q x^{m-1},\quad [m]_q={q^{2m}-1\over q^{2}-1},
\quad \del_q (x^mx^n)=(\del_q x^m)x^n+ q^{2m} x^m(\del_q x^n).}
We see that the generator of braided-addition on the braided line is the usual
$q$-derivative well-known in $q$-analysis\cite{And:ser}. Its well-known
skew-Leibniz rule is seen now to be nothing other than the rule for a
braided-differential where $\del_q$ gives a factor of $q^{2m}$ as it goes past
$x^m$. This factor $q^2$ plays precisely
the role of $\pm 1$ in the super case.

This differentiates the right regular coaction $\und\Delta x=x\tens
1+1\tens\eps$ and is in fact more properly viewed as a derivative acting from
the right. Another option is to use the left regular coaction $\und\Delta
x=\eps\tens 1+1\tens x$ with the same results in this simplest case.

At a primitive root of unity we are in the anyonic situation\cite{Ma:any}.
There we have defined the enveloping algebra of the one-dimensional anyonic Lie
algebra $U_n(\C)$ with one generator $\xi$ and
\eqn{any-line}{\xi^n=0,\quad\Psi(\xi\tens\xi)=e^{2\pi\imath\over n}
\xi\tens\xi,\quad \und\Delta\xi=\xi\tens 1+1\tens \xi.}
This forms a braided-Hopf algebra (an anyonic one) and one can check that
$(\xi+\xi')^n=0$ if we remember the statistics $\xi'\xi=e^{2\pi\imath\over
n}\xi\xi'$ (the anyonic degree is $|\xi|=1$). Differentiating the anyonic
addition law gives $\del_q$ similar to the formula above and obeying a
braided-Leibniz rule with $\del_q$ of anyonic degree -1.
This is
\eqn{bra-leib}{\del_q (ab)=(\del_q a)b+\cdot\Psi^{-1}(\del_q\tens a)b=(\del_q
a)b+e^{2\pi\imath |a|\over n} a(\del_q b)}
where the braiding is computed from (\ref{Psi-any}) with $|\del_q|=-1$. Note
that we must take the degree of $\del_q$ to be -1 so that the map $\del_q\tens
a\to \del_q a$ is degree-preserving (a morphism in the anyonic category). This
then forced us to use $\Psi^{-1}$ rather than $\Psi$.

Mathematically, we can formalise this braided Leibniz rule as the statement
that one copy of $U_n(\C)$ (regarded as the anyonic enveloping algebra of the
one-dimensional Lie algebra) and equipped with the opposite coproduct as in
Lemma~4.6,  acts on another copy of itself (regarded as an anyonic line
co-ordinate algebra) by $\xi\to \del_q$. The anyonic line becomes a left
braided  $U_n(\C)^{\rm cop}$-module algebra in the sense of (\ref{C-modalg})
but living in $\bar{\CC}$ (with the opposite braid crossing to the one shown in
(\ref{C-modalg})). Since 1 is always bosonic, the opposite coproduct
$\Psi^{-1}\circ\und\Delta$ looks just the same as in (\ref{any-line}) but now
extended as a braided-Hopf algebra in $\bar{\CC}$ (so using $\Psi^{-1}$ to
compute the value on products). Putting this into (\ref{C-modalg}) with the
opposite braiding means that $\del_q$ obeys the braided-Leibniz rule as
described.

The generalization of these constructions to the higher-dimensional case is
straightforward using the braided covectors and braided vectors in Chapter~2.
We differentiate their braided-addition law (\ref{bra-vec}) from the left. The
result is
\eqn{diff-covec-left}{\del^i x_j=\delta^i{}_j,\quad \del^i(x_jx_k)=(\del^i
x_j)x_k+x_a R^a{}_j{}^i{}_b (\del^b x_k),\quad {\rm etc.}}
One can check directly that the $\del^i$ themselves obey the algebra relations
of $V(R')$ and that we have a well-defined action $V(R')^{\rm cop}\tens
V^*(R')\to V^*(R')$ making $V^*(R')$ into a left braided-module algebra in
$\bar{\CC}$, as in (\ref{C-modalg}) but with the opposite braiding. Thus the
general Leibniz rule is
\eqn{covec-leib}{\del^i(ab)=(\del^i a)b+\cdot\Psi^{-1}(\del^i\tens a)b}
where the statistics between $\del^i$ and $x_j$ is as between $v^i$ and $x_j$
in (\ref{bra-co-vec-stat}). Here the coproduct on $V(R')^{\rm cop}$ is the
linear one as usual but extended to products with the opposite braid statistics
 (this simply means with $R_{21}^{-1}$ in place of $R$ in (\ref{bra-covec})).
Iterating (\ref{covec-leib}) along the lines of (\ref{diff-covec-left}) leads
to
\eqn{diff-formula}{\del^i(\vecx_1\cdots\vecx_{n})=e^i_1\vecx_2\cdots\vecx_n
(1+(PR)_{12}+(PR)_{12}(PR)_{23}+\cdots+(PR)_{12}\cdots (PR)_{n-1 n})}
where $e^i=\{\delta^i{}_j\}$ is a basis covector.

Thus, the vector fields on a braided-covector space (such as on $q$-Minkowski
space in the last section) obtained from differentiating the braided addition
law, obey the relations of the braided-vectors $V(R')$. This completes the
braided-geometrical picture of Chapter~7.2. There are several variants of these
formulae, including ones where the $\del^i$ act from the right.

These constructions can also be understood as variants of some of our
diagrammatic results in Chapter~4. For any braided-Hopf algebra we can use
Lemma~4.13 applied to $\und\Delta:B\to B\und\tens B$ (the left regular
coaction) to have a left action of $B^{*\rm cop}$, by evaluating the left hand
output of $\und\Delta$ as proven in (\ref{C-B*modalg}) and working in the
category $\bar C$ with reversed braid crossings. The above construction is a
variant of something like this and lives in the category with reversed $\Psi$.
Recall that we could also convert the diagrammatic action over to a right
action of $B^*$ in our original category using the braided-antipode, with
resulting right action (\ref{right-vect}).

The abstract diagrammatic constructions are especially useful when we have good
information about $B^*$. Thus we can apply them to braided matrices and braided
groups to obtain the corresponding left-invariant and right-invariant
braided-vector fields. For example, the canonical right-action of $B^*$ on $B$
in (\ref{right-vect}) in the case when $B=B(A,A)$ (the braided group of
function algebra type in Chapter~5) comes out from Proposition~5.9 and
(\ref{right-vect}) as
\eqn{right-vect-B(A)}{a\ra b=\sum <\CR\ut\la b, a\o>a\t<\CR\uo,a\th>,\quad a\in
B(A,A),\ b\in B(H,H).}
This makes $B(A,A)$ into a right braided $B(H,H)$-module algebra in our
category. For example, on $BSL_q(2)$ as in Chapter~2 there is an action of the
braided-group associated to $U_q(sl_2)$ (which has the same algebra but the
matrix braided-coproduct as explained in Chapter~6.2). Similarly for any
$U_q(g)$ in FRT form. The action computed from (\ref{right-vect-B(A)}) comes
out on generators as
\eqn{right-vec-mat}{u^i{}_j\overleftarrow {\del}^k{}_l=
u^a{}_b \widetilde{R}{}^k{}_n{}^m{}_j Q^n{}_p{}^i{}_a R^b{}_m{}^p{}_l}
and must now obey the right braided-module algebra axioms as in the proof of
Lemma~4.14. Writing the action $\ra L^i{}_j=\ra
(\vecl^+S\vecl^-)^i{}_j=\overleftarrow{\del}{}^i{}_j$, this comes out as
\eqn{mat-leib}{ (ab)\overleftarrow{\del}{}^i{}_j= a\cdot\Psi(b\tens
\overleftarrow{\del}{}^i{}_k)\overleftarrow{\del}{}^k{}_j}
using the braided-matrix coproduct on the braided-group version of $U_q(g)$.
For example, when we compute $\vecu_1\vecu_2\overleftarrow{\del}{}_3$ (in the
compact notation), we already know the braiding of $\vecu_2$ with
$\overleftarrow{\del}{}_3$ since it is the same as between $\vecu,\vecu'$ in
(\ref{bra-mat}) as computed in \cite{Ma:exa}\cite{Ma:eul}. In the compact
notation this means
\eqn{mat-leib-B(R)}{R_{23}^{-1}(\vecu_1
\vecu_2)R_{23}\overleftarrow{\del}{}_3=(\vecu_1\overleftarrow{\del}{}_3)
R_{23}^{-1}(\vecu_2 R_{23}\overleftarrow{\del}{}_3).}

There are as usual plenty of variants of these  results with left-right
interchanged etc, as well the more standard regular actions dualised as in
Lemma~4.13 and working with the variant $B(H,H)^{\rm cop}$ and the opposite
braiding. These left or right regular actions are `matrix-differentials' in the
sense of (\ref{mat-leib}) etc. and define the differential structure on the
braided group (just as the left regular or right regular representation defines
the differential structure on a usual Lie group).
The semidirect product by such actions can be called the braided Weyl algebra.

Of course, every braided module algebra action induces some kind of braided
vector field. For example, we have already seen in Chapter~4 that there is a
braided-adjoint action, so $B(H,H)$ acts on itself from the left by this. As a
linear map it coincides with the usual quantum adjoint action already given in
(\ref{q-ad-act-mat}), giving Ad-vector fields $\del^i{}_j=L^i{}_j\la$ obeying
the braided left-module algebra axioms. On the generators this means
\eqn{mat-lie-left}{\del^i{}_j(ab)=\del^i{}_k\Psi(\del^k{}_j\tens a)b.}
Here the statistics between $\del$ and $L$ are again as braided matrices. So,
for example,
\eqn{mat-leib-B(R)-left}{\del_1 R_{12}(\vecu_2\vecu_3)R_{12}^{-1}=(\del_1
R_{12} \vecu_2) R_{12}^{-1}(\del_1\vecu_3)}
etc.

It would be interesting to compare and contrast these braided-group
constructions to the physicists approach to quantum differentials in
\cite{WesZum:cov}\cite{Zum:int} and elsewhere.

\end{document}